\documentclass[11pt]{article}

\usepackage{graphicx}
\usepackage{amsmath,amssymb,amsthm}
\usepackage[latin1]{inputenc}
\usepackage[T1]{fontenc}
\usepackage{diagbox}
\usepackage{array,multirow}

\usepackage[intoc]{nomencl}
\usepackage{natbib}
\usepackage[french,english]{babel}

\setcitestyle{citesep={~;}}
\setcitestyle{aysep={}}
\let\cite=\citep

\usepackage{lmodern}
\usepackage{dsfont}

\title{Gravitational wave detectors from an experimental perspective}  
 
\author{\textbf{Authors}\\
M.~Trad-Nery, M.~Turconi, W.~Chaibi\\
\footnotesize{Universit\'e C\^ote d'Azur, Observatoire de la C\^ote d'Azur,}\\
\footnotesize{Laboratoire ARTEMIS, CNRS, Bd de l'Observatoire, 06300 Nice, France}}

\makenomenclature 
\makeindex

\begin{document} 

\maketitle
\tableofcontents

\noindent
\textbf{Foreword}

This chapter introduces the fundamental principles of gravitational wave detectors (GWDs) in a simple and comprehensive manner. Because these instruments aim for extremely high sensitivity, it is essential to understand their various noise sources, how such noise couples to the detector output, and the strategies used to mitigate them. We begin with a description of noise from an experimental physicist's perspective, including an introduction to spectral densities, which are widely used to describe the detector sensitivity and noise. 
In the second section, the effect of a gravitational wave on a light field is calculated from the modification of the space-time metric. We then show how a simple Michelson interferometer can be used to detect gravitational waves and which parameters can be modified to enhance the gravitational wave signal. This section also introduces how different noise sources couple into the interferometer readout. 
From the third section onward, we move from a basic Michelson interferometer to a more realistic model of a gravitational-wave detector. For that, we introduce Fabry-Perot cavities and show how they can be used in the interferometer arms to increase the signal of a gravitational wave (GW), and how they are used as a filter for the spatial profile of the laser beam. We also discuss their use as sensors in frequency-stabilization feedback loops, and provide a general introduction to control systems that will be used later in the chapter. The advanced detector configuration is then described: a Michelson interferometer equipped with Fabry-Perot arm cavities, a power recycling mirror, laser power and frequency stabilization, and a mode-cleaner cavity. 
Finally, we focus on the main noise sources that limit detector performance: seismic noise, thermal noise, and quantum noise. This description is made with a simple concept of a harmonic oscillator, described both in the classical and quantum approaches. To conclude, these noises contributions are computed in the frame of the Virgo detector and a sensitivity curve is calculated. Although a simplified layout of a gravitational wave detector is considered, it takes into account the most dominant effects and yields in a sensitivity estimate close to the what is observed in real detectors.

\section{The noise for an experimental physicist}

As for any metrology experiment, setting up a gravitational wave detector requires managing a variety of noise sources of different physical origins. Measuring those noise sources and understanding their coupling paths to the detector is as important as identifying their nature. It requires a deep understanding of the corresponding mathematical basis and its physical interpretation. In the following sections, we present the definition of time dependent noise (relevant to gravitational wave detectors) with the concern of relating it to the measurement process in classical physics.

\subsection{Noise definition}

Noise can be defined as a random (or unpredictable) signal that adds to the physical quantity to be measured, potentially masking or distorting its true value. The total signal $\mathcal{S}$ (which is real) that is measured can be written as: 
\begin{equation}
    \mathcal{S} = x+\epsilon , 
\end{equation}
where $x$ is the quantity to be measured and $\epsilon$ is the noise due to a random process and is characterized by its probability density $p\left(\epsilon\right)$.
In the specific case where the noise is time-dependent, one can write:
\begin{equation}
    \mathcal{S}\left(t\right) = x\left(t\right) + \epsilon\left(t\right) , 
\end{equation}
where for each time $t$, $\epsilon\left(t\right)$ is a time dependent random process characterized by the time-dependent probability density $p_t\left(\epsilon\right)$. 

In order to extract $x\left(t\right)$ out of the measurement $\mathcal{S}\left(t\right)$, one has to address the noise, which will be the focus of the next sections.

\subsection{Noise characterization}
From now on, only time dependent random processes will be considered. They are partially characterized by their corresponding time varying probability density. Hereafter, we give an infinite set of parameters defined for a single time $t$ as an alternative to characterize the process:
\begin{itemize}
    \item The mean value:
\begin{equation}
    \mu_{\epsilon}\left(t\right)=\langle \epsilon(t)\rangle= \int_{-\infty}^{+\infty} \epsilon(t)\times p_t\left(\epsilon\right) \,d\epsilon \label{eq:meanvalue} ,
\end{equation}
    \item The variance:
\begin{equation}
    \mu_{\epsilon,2}(t)=\sigma_\epsilon^2\left(t\right)=\langle \left(\epsilon(t)-\mu_{\epsilon}\left(t\right)\right)^2\rangle= \int_{-\infty}^{+\infty} \left(\epsilon(t)-\mu_{\epsilon}\left(t\right)\right)^2\times p_t\left(\epsilon\right) \,d\epsilon ,
\end{equation}
    \item The moment of order $n$:
\begin{equation}
    \mu_{\epsilon,n}\left(t\right)=\langle \left(\epsilon\left(t\right)-\mu_{\epsilon}\left(t\right)\right)^n\rangle= \int_{-\infty}^{+\infty} \left(\epsilon(t)-\mu_{\epsilon}\left(t\right)\right)^n\times p_t\left(\epsilon\right) \,d\epsilon ,
\end{equation}
 \item For any function $f$ the mean value of $f(\epsilon)$:
\begin{equation}
    \label{eq:fe} 
    \mu_{f(\epsilon)}\left(t\right)=\langle f\left(\epsilon\left(t\right)\right)\rangle= \int_{-\infty}^{+\infty} f\left(\epsilon\right)\times p_t\left(\epsilon\right) \,d\epsilon .
\end{equation} 
\end{itemize}
In the last formulae, the brackets correspond to an average procedure on the outcomes of an ensemble of identical experiments, all evaluated at time $t$. 
Knowing this infinite  set of parameters (infinite moments $n$) fully substitutes the probability density $p_t\left(\epsilon\right)$ for a given time $t$, but it doesn't give any information on how it evolves with time. We will see next that the auto-correlation functions will give this information. 
 
\subsection{Auto-correlation}

We are now interested in characterizing the relationship between $\epsilon(t_1)$ and $\epsilon(t_2)$ which is not deterministic. One can analyze the so-called auto-correlation function $\Gamma_{\epsilon}$, defined as:
\begin{align}
 \Gamma_{\epsilon}\left(t_1,t_2\right)
   & = \int_{-\infty}^{+\infty} \int_{-\infty}^{+\infty} \epsilon_1\times \epsilon_2 \times p_{t_1,t_2}\left(\epsilon_1,\epsilon_2\right)\,d\epsilon_1\,d\epsilon_2 \\     &= \langle\left(\epsilon\left(t_1\right)\epsilon\left(t_2\right)\right) \rangle, \nonumber&&
\end{align}where $p_{t_1,t_2}\left(\epsilon_1,\epsilon_2\right)$ is the joint probability density function and $\epsilon_j = \epsilon (t_j),\,\left(j=1,2\right)$. Additional information can be obtained by analyzing the auto-correlation functions of order $n$ involving $n-2$ intermediate times $t_1',t_2'...t_{n-2}'$ between $t_1$ and $t_2$: 
\begin{eqnarray}
    \nonumber &&\Gamma_{\epsilon}^{\left(n\right)}\left(t_1,t_1',...,t_{n-2}',t_2\right)\\
    \nonumber &=&\int_{-\infty}^{+\infty} \int_{-\infty}^{+\infty}... \epsilon_1\times \epsilon_1' \times...\times \epsilon_2 \times p_{t_1,t_1',...,t_2}\left(\epsilon_1,\epsilon_1',...,\epsilon_2\right)\,d\epsilon_1\,d\epsilon_1'...d\epsilon_2\\
    &=&\langle\left(\epsilon\left(t_1\right)\epsilon\left(t_1'\right)...\epsilon\left(t_{n-2}'\right)\epsilon\left(t_2\right)\right) \rangle .
\end{eqnarray}with $\epsilon_j'=\epsilon\left(t_j'\right)$. In practice, however, the second order auto-correlation function already gives enough information about the considered process.

\subsection{Stationary noise}
\label{Stationary_noise}

\begin{figure} [h!]
\centering \includegraphics[width=1\textwidth]{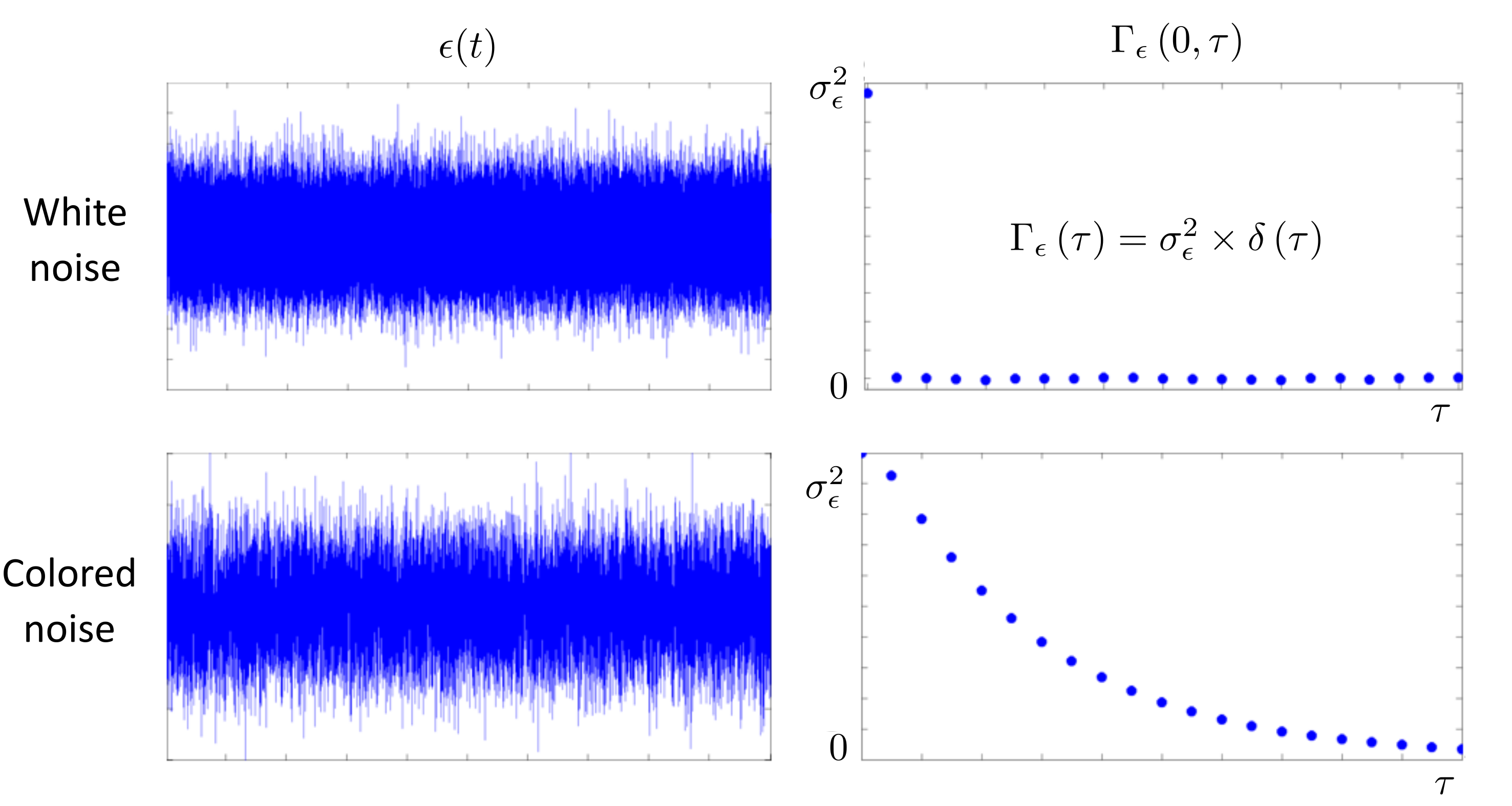}
\caption{Two examples of noises: white and colored. On the temporal profile (left), these two noises are hardly distinguishable, whereas their auto-correlation functions (right) have different behaviors.}
\label{white_colored_noise}
\end{figure}

In the specific case of stationary noise, the noise characteristics (set of parameters given by Equations \ref{eq:meanvalue} to \ref{eq:fe}) do not change with time. Hence, for all sets of times $t_1,t_2,...,t_n$ and any delay $T$, one has:
\begin{equation}
\label{stationary_noise}
    p_{t_1,t_2,...,t_n}\left(\epsilon_1,\epsilon_2,...,\epsilon_n\right)=p_{t_1+T,t_2+T,...,t_n+T}\left(\epsilon_1,\epsilon_2,...,\epsilon_n\right) .
\end{equation}
An interesting case is the 2nd order stationary and centered noise where the condition\,\ref{stationary_noise} stands up to $n=2$: 
\begin{equation}
p_t(\epsilon) = p_{t+\tau}(\epsilon) = p(\epsilon)\\ \textrm{and} \\
p_{t_1,t_2}(\epsilon_1,\epsilon_2) = p_{t_1+\tau,t_2+\tau}(\epsilon_1,\epsilon_2),
\end{equation}and the noise has a zero mean value:
\begin{equation}
    \mu_\epsilon\left(t\right)=\langle\epsilon\left(t\right)\rangle=0.
\end{equation}
The standard deviation is $    \sigma_\epsilon^2\left(t\right)=\sigma_\epsilon^2$,
and the second order auto-correlation only depends on the delay $\tau = t_2-t_1$. We then have :
\begin{equation}
\left\{
    \begin{array}{ll}
        &\Gamma_{\epsilon}\left(t,t+\tau\right)=\Gamma_{\epsilon}\left(0,\tau\right),\\
        \\
        & \sigma_\epsilon^2=\Gamma_\epsilon\left(0,0\right)\geq \Gamma_\epsilon\left(0,\tau\right),
    \end{array}
\right.    
\end{equation}where the inequality can be demonstrated by calculating $\langle\left(\epsilon\left(t\right)\epsilon\left(t+\tau\right)\right)^2\rangle$. It should be noted that the last equation only stands for classical process and is violated in quantum mechanics \cite{Bachor2019}.
On Figure \ref{white_colored_noise}, two types of noise characterized by different behaviors of the auto-correlation functions are shown: white noise for which the correlation between two different times is strictly equal to zero, i.e. $\Gamma_\epsilon\left(\tau\right)=\sigma_\epsilon^2\times\delta\left(\tau\right)$, and colored noise for which the auto-correlation function varies smoothly from its maximum down to zero.

\subsection{Temporal average and ergodicity}
\label{Temporal_average_ergodicity}

Let us examine the situation in which only a single experiment is available. For any function $f$ applied on the noise process $\epsilon\left(t\right)$, one can define the temporal average $m_f$:
\begin{equation}
    m_f=\overline{f\left(\epsilon\right)}=\lim_{T\to\infty}\frac{1}{2T}\int_{-T}^{T} f\left(\epsilon\left(t\right)\right)\,dt ,
\end{equation}with the assumption that this limit actually exists. In particular, one can define the mean value, the standard deviation, and the auto-correlation function of the noise $\epsilon$ based on the temporal average:
\begin{equation}
    m_\epsilon=\overline{\epsilon}=\lim_{T\to\infty}\frac{1}{2T}\int_{-T}^{T} \epsilon\left(t\right)\,dt ,
\end{equation}
\begin{equation}
\label{variance_temporal}
    s_\epsilon^2=\overline{\epsilon^2}=\lim_{T\to\infty}\frac{1}{2T}\int_{-T}^{T} \epsilon^2\left(t\right)\,dt ,
\end{equation}
\begin{equation}
    C_\epsilon\left(\tau\right)=\overline{\epsilon\left(t\right)\epsilon\left(t+\tau\right)}=\lim_{T\to\infty}\frac{1}{2T}\int_{-T}^{T} \epsilon\left(t\right)\epsilon\left(t+\tau\right)\,dt , 
\end{equation}
\begin{eqnarray}
    \nonumber C_\epsilon^{\left(n\right)}\left(\tau_1,...,\tau_{n}\right)&=&\overline{\epsilon\left(t\right)\epsilon\left(t+\tau_1\right)...\epsilon\left(t+\tau_{n}\right)}\\
    &=&\lim_{T\to\infty}\frac{1}{2T}\int_{-T}^{T} \epsilon\left(t\right)\epsilon\left(t+\tau\right)...\epsilon\left(t+\tau_{n}\right)\,dt . 
\end{eqnarray}
Again, this is considered for a single experiment. In the case of multiple similar experiments in which all temporal averages remain quantitatively the same, the process is called ergodic. The specific case of 2nd order ergodicity:
\begin{equation}
    m_{\epsilon}^{(i)}=m_{\epsilon}^{(j)}=m_{\epsilon}=\overline{\epsilon},
\end{equation}
\begin{equation}
    s_{\epsilon}^{(i)2}=s_{\epsilon}^{(j)2}=s_{\epsilon}^2=\overline{\epsilon^2},
\end{equation}
\begin{equation}
    C_{\epsilon}^{(i)}\left(\tau\right)=C_{\epsilon}^{(j)}\left(\tau\right)=C_{\epsilon}\left(\tau\right)=\overline{\epsilon\left(t\right)\epsilon\left(t+\tau\right)},
\end{equation}is of particular interest for the rest of this section ($(i)$ and $(j)$ label two different experiments).

\subsection{Stationary and ergodic process}

The main class of process one can deal with in experimental physics is the so called (2nd order) stationary and (2nd order) ergodic, for which both conditions described in sections \ref{Stationary_noise} and \ref{Temporal_average_ergodicity} are verified. Additionally, we'll consider this process to be centered (zero mean) as it is the case for most of noises $\epsilon\left(t\right)$. For a given fuction $f\left(\epsilon\right)$, one can use either averages over different experiments or over time:
\begin{equation}
    \mu_{f\left(\epsilon\right)}\left(t\right)=\langle f\left(\epsilon\left(t\right)\right)\rangle = \int_{-\infty}^{+\infty}f\left(\epsilon\right)\times p_t\left(\epsilon\right)\,d\epsilon,
\end{equation}
\begin{equation}
    m_{f\left(\epsilon\right)}=\overline{f\left(\epsilon\right)}=\lim_{T\to\infty}\frac{1}{2T}\int_{-T}^{T}f\left(\epsilon\left(t\right)\right)\,dt.
\end{equation}Since $\epsilon$ is stationary and ergodic these two averages will be equal:
\begin{equation}
    \begin{array}{lllll}
        \mu_{f\left(\epsilon\right)} & = & \overline{\mu_{f\left(\epsilon\right)}\left(t\right)} & \mbox{stationarity}\\
        \\
        &=&\overline{\langle f\left(\epsilon^{(i)}\left(t\right)\right)\rangle}& \\
        \\
        &=&\langle \,\overline{f\left(\epsilon^{(i)}\left(t\right)\right)}\,\rangle&\mbox{linearity}\\
        \\
        &=& \langle m_{f\left(\epsilon^{(i)}\right)}\rangle\\
        \\
        &=&\langle m_{f\left(\epsilon\right)}\rangle & \mbox{ergodicity}\\
        \\
        &=& m_{f\left(\epsilon\right)}.
        \end{array}
\end{equation} More specifically:
\begin{equation}
    m_\epsilon = \mu_\epsilon=0\hspace{1cm}\textrm{;}\hspace{1cm}s_\epsilon^2 = \sigma_\epsilon ^2\hspace{1cm}\textrm{;}\hspace{1cm} C_\epsilon\left(\tau\right) = \Gamma_\epsilon\left(\tau\right) .
\end{equation} From now on, all noises will be considered as (2nd order) stationary and ergodic, and averages will be, case by case, computed following either of the two procedures. 

\subsection{Harmonic analysis: The Fourier transform}

The Fourier transform is the optimal tool to perform harmonic analysis. We'll see in the following that depending on the type of process, it is applied differently with different physical interpretations. 
\subsubsection{Finite energy process}
From a mathematical point of view, a finite energy ($\mathcal{E}$) real process satisfies the condition:
\begin{equation}
\mathcal{E} = \int_{-\infty}^{+\infty}\epsilon\left(t\right)^2\,dt, \end{equation}which guarantees the existence of its Fourier transform $\tilde{\epsilon}$:
\begin{equation}
    \tilde{\epsilon}\left(f\right)=\int_{-\infty}^{+\infty}e^{-2i\pi f t}\epsilon\left(t\right)\,dt .
\end{equation}The Parseval equality relates the energy $\mathcal{E}$ to the Fourier transform $ \tilde{\epsilon}\left(f\right)$ as:
\begin{equation}
    \mathcal{E}=\int_{-\infty}^{+\infty}\left|\tilde{\epsilon}\left(f\right)\right|^2\,df .
\end{equation}Hence, $\left|\tilde{\epsilon}\left(f\right)\right|^2$ represents the energy density per spectral interval $df$ and is called the Energy Spectral Density (ESD).
\subsubsection{Continuous process}
By definition, a continuous process $\epsilon\left(t\right)$ has an infinite energy since it lasts indefinitely (example: the energy of a continuous wave (cw) laser beam):
\begin{equation}
    \int_{-\infty}^{+\infty}\epsilon\left(t\right)^2\,dt = \infty .
\end{equation}However, it is expected to have a finite mean power:
\begin{equation}
\label{mean_power_process}
    \overline{P_\epsilon}=\lim_{T\to+\infty}\frac{1}{2T}\int_{-T}^{T}\epsilon\left(t\right)^2\,dt . 
\end{equation}
For the case of a stationary and ergodic noise and according to Equation \ref{variance_temporal} and Equation \ref{mean_power_process}, the noise mean power is directly given by the variance: $ \overline{P_\epsilon}=\sigma_\epsilon^2$ . 

If it exists, we are interested in the quantity:
\begin{equation}
\label{definition_PSD_limit}
    S_\epsilon\left(f\right)= \lim_{T\to+\infty}\frac{1}{2T}\left|\tilde{\epsilon}_T\left(f\right)\right|^2=\lim_{T\to+\infty}\frac{1}{2T}\left|\int_{-T}^{T}e^{-i2\pi f t}\epsilon\left(t\right)\,dt\right|^2 , 
\end{equation}where $\tilde{\epsilon}_T$ is the Fourier transform of the truncated function $\epsilon\left(t\right)$ over $\left[-T,T\right]$. Summed over all frequencies $f$ one obtains:
\begin{equation}
\label{PSD_variance}
    \begin{array}{lll}
    \int_{-\infty}^{+\infty} S_\epsilon\left(f\right)\,df&=&\lim_{T\to+\infty}\frac{1}{2T}\int_{-\infty}^{+\infty}\left|\tilde{\epsilon}_T\left(f\right)\right|^2\,df\\
    \\
     &=&\lim_{T\to +\infty}\frac{1}{2T}\int_{-\infty}^{+\infty}\epsilon_T\left(t\right)^2\,dt\\
    \\
    &=&\lim_{T\to +\infty}\frac{1}{2T}\int_{-T}^{+T}\epsilon\left(t\right)^2\,dt\\
    \\
    &=&\sigma_\epsilon^2 = \overline{P_\epsilon} . 
    \end{array}    
\end{equation}Therefore, if $S_\epsilon\left(f\right)$ exists, it represents the power density per spectral interval $df$ and is then called Power Spectral Density (PSD).

\subsubsection{The Wiener-Khintchine Theorem}

This theorem states that for a stationary and ergodic process (like the type of noise we are dealing with) the limit defining the PSD $S_\epsilon\left(f\right)$ (Equation \ref{definition_PSD_limit}) does exist and it is given by the Fourier transform of the corresponding auto-correlation function:
\begin{eqnarray}
    \nonumber S_\epsilon\left(f\right)&=&\lim_{T\to\infty}\frac{1}{2T}\left|\int_{-T}^{T}e^{-i2\pi f t}\epsilon\left(t\right)dt\right|^2\\
    &=&\int_{-\infty}^{+\infty}e^{-2i\pi f \tau}\,\Gamma_\epsilon\left(\tau\right)\,d\tau .
\end{eqnarray}
The auto-correlation form of the PSD is very useful for analytic calculations, whereas for experimental measurements, we limit the integration in Equation \ref{definition_PSD_limit} on a properly chosen acquisition time $T_\text{acq}$ and we average it over a number $N$ of successive measurements:
\begin{equation}
S_\epsilon \left(\nu\right) \simeq \frac{1}{T_\text{acq}}\langle\left|\int_{0}^{T_\text{acq}} e^{-i2\pi f t}\epsilon(t)dt\right|^2\rangle _N .
\end{equation}In practice, $T_\text{acq} \gg 1/f_\text{min}$ where $f_\text{min}$ is the minimum frequency we might be interested in.
It is usual to characterize the noise by its Amplitude Spectral Density (ASD) which is the square root of its PSD. It results in an usual unit, which can be seen on the simple case of a voltage noise $\delta V(t)$:  
\begin{align}
\nonumber & \delta V(t) \rightarrow \text{V}, \\
\nonumber & \Gamma_{\delta V}(\tau) \rightarrow \text{V}^2, \\
\nonumber & S_{\delta V}(f) \rightarrow \text{V}^2/\text{Hz},\\
\nonumber & \sqrt{S_{\delta V}(f)} \rightarrow \text{V}/\sqrt{\text{Hz}}.
\end{align}


\section{Detection principle of an interferometric ground based gravitational wave detector}

Ground-based gravitational-wave detectors employ a Michelson interferometer configuration that transduces the gravitational-wave-induced phase shift in the interferometer arms into a measurable change in optical power at its output. In this section, we will compute the output power of the interferometer as a function of a gravitational wave (GW) signal and of usual noise sources. We begin by providing a simplified description of the gravitational wave arriving at the detector, which will then allow us to compute its effect on the light propagating in the interferometer.

\subsection{Gravitational wave effect on a light field}

\begin{figure} [h!]
\centering \includegraphics[width=1\textwidth]{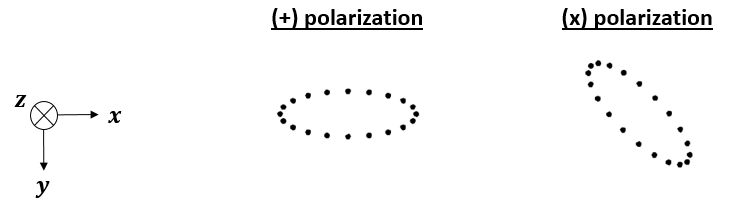} 
\caption{Usual representation of the GW polarizations}
\label{GW_polarizations}
\end{figure}
Far from the source, the GW can be considered as a plane wave. For the sake of simplicity, we will consider its propagation along the $z$ axis orthogonal to the plane $\left(x,y\right)$ in which the light propagates (see Figure \ref{GW_polarizations}). Its polarization in the plane $\left(x,y\right)$ can be projected on the set of orthogonal polarizations $\left(+\right)$ and $\left(\times\right)$. As represented in Figure \ref{GW_polarizations}, the $\left(+\right)$ polarization corresponds to the elongation of the space time along one axis (here $x$) and the $\left(\times\right)$ polarization corresponds to a similar elongation at 45$^{\circ}$. 
Mathematically speaking, the GW is described by its modification of the local space-time metric:
\begin{equation}
g_{\mu\nu} = \eta_{\mu\nu} + h_{\mu\nu},
\end{equation}where $\eta_{\mu\nu}$ is the Minkowski metric describing a flat space time (without perturbation): 
\begin{equation}
\eta_{\mu\nu} = \begin{pmatrix} 
1 & 0 & 0 & 0 \\
0 & -1 & 0 & 0 \\
0 & 0 & -1 & 0 \\
0 & 0 & 0 & -1
\end{pmatrix}.
\end{equation}In the TT (Transverse Traceless) gauge\cite{Reitze2019}, the GW is described by the time dependent perturbation $h_{\mu\nu}$ which we simplified to a pure $\left(+\right)$ cross polarization for the sake of simplicity:

\begin{equation}
h_{\mu\nu} = \begin{pmatrix}
0 & 0 & 0 & 0 \\
0 & +h(t) & 0 & 0 \\
0 & 0 & -h(t) & 0 \\
0 & 0 & 0 & 0
\end{pmatrix}.
\end{equation}

By definition, light follows the geodesic equation, which in Einstein's notations, is written as:
\begin{equation}
\label{geodesic}
g_{\mu\nu}dx^\mu dx^\nu = 0,
\end{equation}where $dx^{\mu(\nu)} = (cdt, dx, dy, dz)$. Expanding Equation \ref{geodesic} gives:
\begin{equation}
c^2dt^2 + dx^2(-1+h(t)) + dy^2(-1-h(t)) - dz^2 = 0 .
\end{equation}
Now, we consider a "classical" photon (as it was defined by Einstein) propagating along the $x$ axis, i.e. $dy = 0$ and $dz = 0$, one obtains:
\begin{equation}
dx = \pm c\sqrt{\frac{1}{1-h(t)}}dt \simeq \pm c\left(1 + \frac{1}{2}h(t)\right)dt .
\end{equation}
%
We now consider two coordinates points 0 and $L$. The coordinates of these two points remains unchanged in the TT gauge. However, if a photon is launched from $x=0$ at $t_0$, it will arrive at $x=L$ at a time $t_L$ that satisfies:
\begin{equation}
L = \int_0^L dx = c\int_{t_0}^{t_L}\left(1 + \frac{1}{2}h(t)\right)dt = c\left(\int_{t_0}^{t_L}dt + \frac{1}{2}\int_{t_0}^{t_L}h(t)dt\right) .
\end{equation}In case the characteristic time of $h\left(t\right)$ is much longer than the unperturbed propagation time $L/c$, $h\left(t\right)$ can be taken out of the integral, leading to: 
\begin{equation}
t_L \simeq t_0 + \frac{L}{c}\left(1 - \frac{1}{2}h(t)\right)\hspace{1cm}\textrm{for the $x$ axis}\label{eq:tx_GWplus},
\end{equation}showing that the GW produces a time delay on the propagation of the photon. The same effect can be computed for the $y$ direction but with an opposite delay:   
\begin{equation}
t_L \simeq t_0 + \frac{L}{c}\left(1 + \frac{1}{2}h(t)\right)\hspace{1cm}\textrm{for the $y$ axis},
\end{equation}which is characteristic of (+) polarization.\\
Let us apply these results obtained for a single photon to the more general case of an electromagnetic field. For a light field propagating along $x$, $E(x,t)$, one can write:
\begin{eqnarray}
\nonumber && \textrm{without GW :}\hspace{0.5cm}  E\left(L,t\right) = E\left(0, t-\frac{L}{c}\right),\\
\nonumber&&\textrm{with GW :}\hspace{0.5cm} E\left(L,t\right) = E\left(0, t-\frac{L}{c}\left(1-\frac{h\left(t\right)}{2}\right)\right)=E\left(L\left(1-\frac{h\left(t\right)}{2}\right), t\right).\\
\end{eqnarray}
So, without GW, the electric field that arrives at a time $t$ at $x=L$ is the electric field that was at $x=0$ at time $t-L/c$. The GW causes a change in the propagation time according to Equation \ref{eq:tx_GWplus} that can also be expressed as a change in the propagation length.\\
The complex electric field that describes a single frequency laser beam at $x=L$ in presence of a GW can be written as:
\begin{equation}
E_L\left(t\right) = E_0\,e^{-i2\pi\nu_0 t}\cdot e^{i\frac{2\pi\nu_0 L}{c}\left(1-\frac{h\left(t\right)}{2}\right)},
\end{equation}
where $\nu_0$ is the laser frequency and $E_0$ the field amplitude.
This equation shows that the GW acts on the phase of a light beam, and since the phase cannot be directly measured with a photodetector, interferometry is an ideal technique to measure gravitational waves.

\subsection{The Michelson interferometer, its output and tuning}
\label{MI}

The detection of GWs using optical interferometry was first proposed in 1963 by Gertsenshtein and Pustovoit \cite{Gertsenshtein1963}. They proposed using a Michelson interferometer as a detector, since it measures the phase difference between two orthogonal arms. This makes it well suited for detecting a GW when the wave's polarization axes are aligned with the interferometer arms.  
In the following sections we compute the output power of the interferometer as a function of the gravitational wave signal, and also as a function of the main noise sources in GWDs.
%
%

\begin{figure}[h!]
\centering \includegraphics[width=1\textwidth]{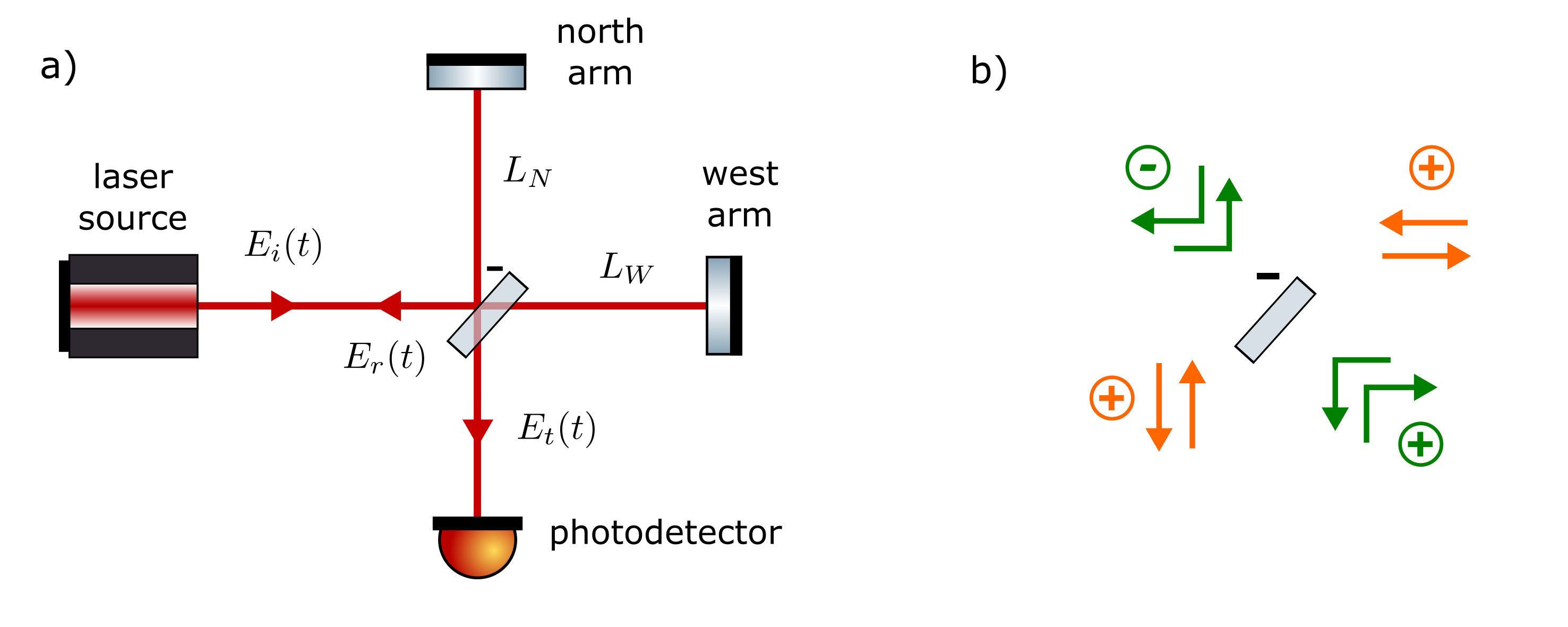} 
\caption{a) Schematics of a Michelson interferometer. Light coming from a laser source is split by a beamsplitter into two orthogonal arms and fully reflected back to the beamsplitter. Depending on the interference, part of the light is reflected back to the laser and part is transmitted to a photodetector. b) Fields in reflection (green arrows) and transmission (orange arrows) of a beamsplitter represented by the real convention, where the reflected light field acquires a minus sign in one (here the left) of the surfaces of the beamsplitter.}
\label{fig:michelson_ifo}
\end{figure}

We begin by calculating the complex electric fields in transmission ($E_t$) and reflection ($E_r$)  of a simple Michelson interferometer (see Figure \ref{fig:michelson_ifo}a) as a function of the complex input field $E_i(t) = E_0 \cdot e^{-i(2\pi\nu_0t+\phi_L(t))}$ and of the round-trip phase $\phi_{N(W)}$ acquired by the fields in the north(west) arm (north an west are the directions of the Virgo interferometer's arms). Here $\phi_L(t)$ is the time dependent phase fluctuation of the input field, which represents the phase noise of the laser source.
%
For this calculation the so called real beamsplitter convention, illustrated on Figure \ref{fig:michelson_ifo}b, will be used. In this convention, the complex field acquires a minus sign in reflection when it impinges on one of the surfaces of the beamsplitter, whereas it remains unchanged when reflected on the other surface. It also remains unchanged for both transmissions. A phase convention is required in order to satisfy the energy conservation condition.
For a 50:50 beamsplitter, like in the case of GWDs, the reflection ($\sqrt{R}$) and transmission ($\sqrt{T}$) field amplitude coefficients are equal to $\sqrt{R}=\sqrt{T}=1/\sqrt{2}$. The transmitted field is the sum of the field transmitted into the west arm that, after a round-trip, is reflected towards the photodetector, and of the field reflected to the north arm that is subsequently transmitted by the beamsplitter. A similar logic applies to the reflected interferometer field. This will lead to the following equations: 
\begin{align}
 E_t(t) & = E_0e^{-i2\pi\nu_0t} \cdot \sqrt{T} \cdot \sqrt{R} \cdot e^{i \phi_W(t)}- E_0e^{-i2\pi\nu_0t} \cdot \sqrt{R} \cdot \sqrt{T} \cdot e^{i \phi_N(t)} \nonumber \\ 
& = \frac{1}{2}E_0e^{-i2\pi\nu_0t}(e^{i\phi_W(t)} - e^{i\phi_N(t)}), \\
 E_r(t) &  = E_0e^{-i2\pi\nu_0t}\cdot T \cdot e^{i \phi_W(t)} + E_0e^{-i2\pi\nu_0t} \cdot R \cdot e^{i \phi_N(t)} \nonumber \\ 
& = \frac{1}{2}E_0e^{-i2\pi\nu_0t}(e^{i\phi_W(t)} + e^{i\phi_N(t)}),
\end{align}
with the round-trip phase on the interferometer arms given by:
\begin{align}
\phi_W = \frac{4\pi\nu_0\cdot L_W}{c} \cdot \left(1 - \frac{h(t)}{2}\right) - \phi_L\left(t - 2L_W/c\right), \\
\phi_N = \frac{4\pi\nu_0\cdot L_N}{c} \cdot \left(1 + \frac{h(t)}{2}\right) - \phi_L\left(t - 2L_N/c\right).
\end{align}
Here $L_{N(W)}$ are the geometrical lengths of the interferometer arms in the absence of a gravitational wave. Note that, to first order, the time argument for $\phi_L$ takes into account only the travel time of the light without the gravitational wave effect. It is useful to re-write these equations as a function of the common ($\bar{\phi}$, $\bar{L}$) and differential ($\Delta \phi$, $\Delta L$) phase and arm lengths, given by:
\begin{align}
\bar{\phi} = \frac{\phi_W + \phi_N}{2} \qquad \text{and} \qquad  \Delta\phi = \frac{\phi_W - \phi_N}{2},\label{eq:common_differential_phases}  \\
\bar{L} = \frac{L_W + L_N}{2}  \qquad \text{and} \qquad  \Delta L = \frac{L_W - L_N}{2}.
\end{align}
This will lead to:
\begin{align}
E_t(t) = iE_0e^{-i(2\pi\nu_0t-\bar{\phi})} \cdot \sin(\Delta\phi), \\
E_r(t) = E_0e^{-i(2\pi\nu_0t-\bar{\phi})} \cdot \cos(\Delta\phi).
\label{eq:refl_ITF}
\end{align}
Since the power is proportional to the modulus square of the electric field, it will only depend on the differential phase $\Delta\phi$. For frequencies $f\ll c/(2\bar{L})$, we obtain:
\begin{align}
\Delta\phi \simeq \frac{4\pi\nu_0\Delta L}{c} - \frac{4\pi\nu_0\bar{L}}{c} \cdot \frac{h(t)}{2} - \frac{4\pi\delta\nu(t)\Delta L}{c}.
\label{eq:deltaphi_mich}
\end{align}
The term $\delta\nu(t)$ is the frequency noise of the laser which has the following relation with the phase noise:
\begin{equation}
    \delta\nu(t) = \frac{1}{2\pi}\frac{d\phi_L(t)}{dt} .
    \label{eq:frequency_noise}
\end{equation}
The transmitted power by the interferometer as a function of the input power $P_0$ is:
\begin{align}
 P_t(t) & = K\times|E_t(t)|^2 \nonumber \\ &= P_0 \cdot \sin^2\left(\frac{4\pi\Delta L}{\lambda_0} - \frac{2\pi \bar{L} }{\lambda_0} \cdot h(t) - \frac{4\pi\delta\nu(t)}{\nu_0}\frac{\Delta L}{\lambda_0}\right).
 \label{eq:mi_pt_1}
\end{align}
where $K$ is a constant depending on the beam shape that will be defined later. One can see that, in the absence of a GW wave, the mean laser power transmitted by the interferometer will be zero when $\Delta L=0$. This operational point is called dark fringe, since it corresponds to a specific destructive interference in transmission (the full mean power is reflected back to the laser source) which is insensitive to the frequency noise . The response for a GW, however, is not linear at this operational point since $P_t$ is quadratic with $h(t)$. For this and other reasons, GW detectors operate close to the dark fringe, by introducing a small arm length offset $\Delta L_\text{DC}$ on a technique called DC readout. With this offset, a small amount of mean power is transmitted by the interferometer, even in the absence of a GW. For this operational point, and considering that $\bar{L} \cdot h(t)/\lambda_0 \ll \Delta L_\text{DC}/\lambda_0 \ll 1$,  Equation \ref{eq:mi_pt_1} can be approximated to:
\begin{equation}
 P_t(t) \simeq \frac{16 \pi^2 P_0 }{\lambda_0^2} \cdot \left(\Delta L_\text{DC}^2 - \bar{L} \cdot \Delta L_\text{DC} \cdot h(t) - \frac{2 \Delta L_\text{DC}^2 \cdot \delta\nu(t)}{\nu_0}\right),
 \label{eq:mi_pt_2}
\end{equation}
which is linear with $h(t)$. This equation shows that the power variation due to the gravitational wave signal is increased by increasing the mean arm length of the interferometer and the laser power. For this reason, ground based GWDs have armlengths of kilometers, and several Watts at the interferometer input. The first term on this equation is the mean transmitted power impinging on the photodetector $\bar{P}_t=16 \pi^2 \Delta L_\text{DC}^2 \bar{P_0}/\lambda_0^2$.

Let us now analyze noise contributions to the transmitted power $P_t(t)$, which will compete with the GW signal that one wants to measure. Already expressed on the Equation \ref{eq:mi_pt_2} is the laser frequency noise, whose coupling to the output power is reduced by reducing the interferometer arm mismatch. 

Another source of noise is laser power noise, which can be split into two type of sources: technical power fluctuations, which will be represented by $\delta P_0(t)$, and fundamental power fluctuations (shot noise), represented by $\delta P_\text{SN}(t)$. Technical power fluctuations originates from different "classical" sources (coupled mechanical noise, electronic noise, and others)  and its coupling can be obtained by substituting $P_0 = \bar{P_0} + \delta P_0(t)$ in Equation \ref{eq:mi_pt_2}. Shot noise is a fundamental noise source originated from the quantum nature of light, which will be tackled in Section \ref{Quantum_noise}. Here we will instead introduce its classical approach. An ideal laser with no technical noise will emit a number of photons $N(t)$ during a time $\Delta t$ that can be detected randomly on a photodetector with a probability following a Poissonian law, i.e., with mean value $\bar{N}$ and variance $\sigma_N^2 = \bar{N}$. Let us now determine the PSD of the corresponding power fluctuations due to shot noise. Over $\Delta t$, the laser power is $P_{\Delta t}=h_p\nu_0 N(t)/\Delta t$ and has a mean value $\bar{P}_{\Delta t}=h_p\nu_0 \bar{N}/\Delta t$, with $h_p$ being the Planck's constant. Following the Poissonian law, the variance of $P_{\Delta t}$ is related to $\sigma_N^2$ by:
\begin{equation}
   \sigma_{P_{\Delta t}}^2 = \left( \frac{h_p\nu_0}{\Delta t} \right)^2 \sigma_N^2=\left( \frac{h_p\nu_0}{\Delta t} \right)^2 \bar{N} =\frac{h_p\nu_0}{\Delta t}\bar{P}_{\Delta t}= \int_{1/\Delta t} h_p\nu_0\bar{P}_{\Delta t}\,df .
\end{equation}The last equality shows that the PSD of the shot noise is equal to:
\begin{equation}
 \label{eq:shotnoise}
    S_\text{SN}(f) = h_p\nu_0\bar{P}_{\Delta t},
\end{equation}according to Equation \ref{PSD_variance}, and that is proportional to the mean laser power. We will give in Section \ref{Quantum_noise} a more realistic quantum model of light and infer the same expression.

As an example, for a typical detected mean power of 10 mW and for a wavelength of $\lambda_0 = 1064\,\text{nm}$, one finds the ASD for the shot noise equal to $4 \times 10^{-11}\,\text{W}\cdot \text{Hz}^{-1/2}$ and a relative shot noise (shot noise divided by the mean laser power) of $4 \times 10^{-9}\,\text{Hz}^{-1/2}$. 

We now get back to the interferometer transmitted power. To calculate the PSD of the shot noise in transmission of the interferometer, one can simply substitute the transmitted mean power, which is given by the first term in Equation \ref{eq:mi_pt_2}, into Equation \ref{eq:shotnoise}.

Finally, any noise source that causes an unwanted differential motion $\delta L_-(t)$ of the interferometer mirrors will couple in transmission of the interferometer and cannot be distinguished from a GW signal. This differential motion could be induced by seismic noise, vibrational noise, or thermal noise, for example. This effect can be taken into account by making the substitution in Equation \ref{eq:mi_pt_2} $(\Delta L_\text{DC})^2 \rightarrow (\Delta L_\text{DC} + \delta L_-(t))^2 \simeq \Delta L_\text{DC}^2 + 2 \Delta L_\text{DC} \cdot \delta L_-(t)$, for small $ \delta L_-(t)\ll \Delta L_{\textrm{DC}}$. Taking into account all discussed noise contributions, Equation \ref{eq:mi_pt_2} can be re-written as:
%
%
\begin{align}
P_t(t) \simeq 
\underbrace{\frac{16 \pi^2 \Delta L_{\mathrm{DC}}^2}{\lambda_0^2} \, \bar{P}_0}_{\text{DC gain factor}} 
\cdot \Bigg(
& 1
+ \underbrace{\frac{\delta P_0(t)}{\bar{P}_0}}_{\text{technical power noise}} 
+ \underbrace{\frac{2 \, \delta L_{-}(t)}{\Delta L_{\mathrm{DC}}}}_{\text{displacement noise}} 
- \underbrace{\frac{\bar{L} \cdot h(t)}{\Delta L_{\mathrm{DC}}}}_{\text{GW signal}} \notag \\
& \quad
- \underbrace{ \frac{2\delta \nu(t)}{\nu_0}}_{\text{frequency noise}}
\Bigg)
+ \underbrace{\delta P_{\mathrm{SN}}(t)}_{\text{shot noise}} .
\label{eq:mi_pt_3}
\end{align}

The signal to noise ratio can be obtained by dividing the PSD of the GW signal and the PSD of the considered noise sources. This leads to:
\begin{eqnarray}
\label{SNR}
\nonumber \rho^2(f) &=& \frac{\bar{L}^2 \cdot S_h(f)}{\Delta L_\text{DC}^2 S_\text{rpn}(f) +   4 S_{\delta L-}(f)+ 4 \Delta L_\text{DC}^2 \cdot S_{\nu}(f) / \nu_0^2 + \Delta L_\text{DC}^2 h_p \nu_0 / \bar{P_t} }, \\
\end{eqnarray}
where $S_{\textrm{rpn}}(f)$ is the PSD of $\delta P_0 (t) / \bar{P_0}$ and represents the relative power noise. In practice one has to choose the adequate armlength offset $\Delta L_{\textrm{DC}}$ taking into account several factors. For our simple approach, we shall consider the sensitivity to be limited by the shot noise, which is actually true in practice for frequencies higher than few hundreds of Hz. The signal-to-shot-noise ratio is then given by:
\begin{equation}
    \rho_{\textrm{SN}}^2\left(f\right)=\frac{16\pi^2\bar{P}_0\bar{L}^2}{h_p c\lambda_0} S_h(f). 
\end{equation}
One can notice that it does not depend on the arm length offset $\Delta L_{\textrm{DC}}$ provided the latter remains small enough to keep the used approximation valid. In a realistic situation, other criteria have to be considered. For example, the transmitted power should be high enough such that the detection is not limited by the photodetector dark (electronic) noise and, at the same time, the detected power should remain below the photodetector saturation power. Further constraints are to be considered related to the defects appearing in real detectors (asymmetries, high power effects, etc).
%
 
\section{Sensitivity enhancement with Fabry-Perot cavities}

\begin{figure}[h]
		\centering
		\includegraphics[width=1.1\textwidth]{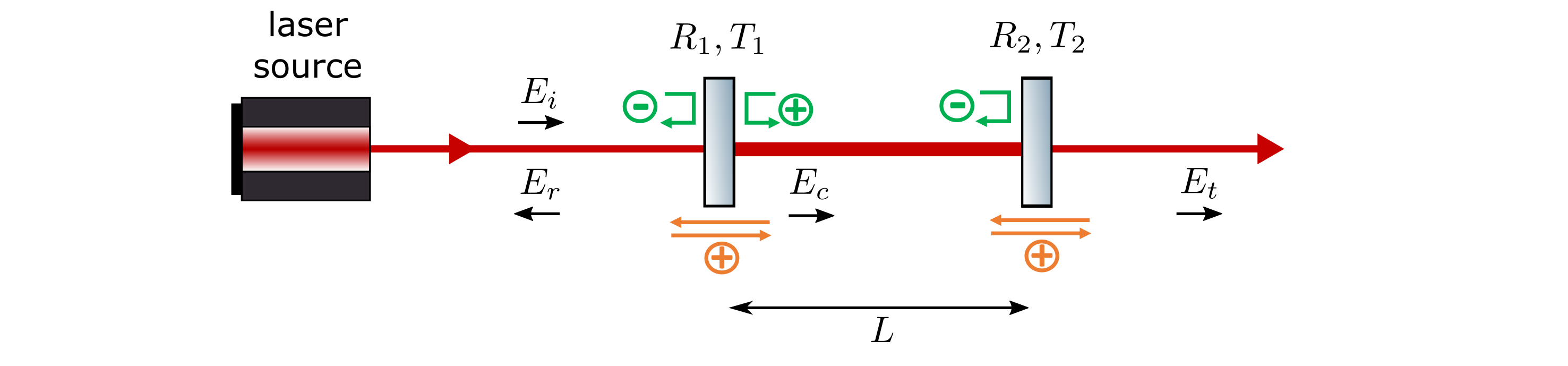}
		\caption{Representation of a Fabry-Perot cavity. The mirrors are characterized by their transmission and reflection coefficients $\sqrt{T_1},\sqrt{R_1}$ and $\sqrt{T_2}, \sqrt{R_2}$ and they are placed at a distance $L$. A polarized laser is injected in the cavity. $E_i, E_r, E_c$ and $E_t$ are respectively the input, reflected, circulating and transmitted complex electric fields.} 
		\label{fig:FP}
	\end{figure}

We have shown that the phase shift induced by a GW in a Michelson interferometer increases with the interferometer's mean arm length $\bar{L}$ (see Equation \ref{eq:deltaphi_mich}). In this section, we discuss how the GW signal can be further enhanced by introducing Fabry-Perot cavities in the interferometer arms. A Fabry-Perot cavity is composed of two mirrors placed face-to-face, like shown in Figure \ref{fig:FP}. An intuitive explanation why they can enhance the detector sensitivity is that the light is "trapped" in the cavity for a certain amount of time, making several round trips between the mirrors, and during which it interacts with the GW. This effectively increases the interferometer's arm length and thus amplifies the GW signal.

Let us now describe the Fabry-Perot behavior, first by considering a linear cavity injected by a laser beam such as the one depicted in Figure \ref{fig:FP}. The mirrors are considered infinitesimally thin and partially reflective with amplitude transmission and reflection coefficients $\sqrt{T_{1(2)}},\sqrt{R_{1(2)}}$, and with the sign convention explained in Figure \ref{fig:michelson_ifo} b. We assume that the mirrors have reflection coefficients close to 1 and no absorption or scattering losses. Therefore the energy conservation laws at the mirrors' surfaces can be written as:     
\begin{equation}
  \begin{cases} 
 T_1 + R_1 = 1 \\
T_2 + R_2 = 1   
\end{cases}
\end{equation}
By using the symbols and the sign convention represented in Figure \ref{fig:FP}, one can write the equations for the circulating and reflected time-dependent electric fields, $E_c (t)$ and $E_r(t)$, at the first mirror position and the equation of the transmitted field $E_t(t)$ at the second mirror position. For example, the circulating field results from the sum of the input field $E_i(t)$ transmitted by the first mirror and the circulating field itself after it traveled a round-trip  length of $2L$. Following this logic, one obtains: 
\begin{align}
E_c(t) &= \sqrt{T_1}E_i(t) - \sqrt{R_2R_1}E_c\left(t - \frac{2L}{c}\right) \label{eq:Ec},\\
E_t(t) &= \sqrt{T_2}E_c\left(t - \frac{L}{c}\right) \label{eq:Et} , \\
E_r(t) &= -\sqrt{R_1}E_i(t) - \sqrt{R_2T_1}E_c\left(t - \frac{2L}{c}\right) .
\end{align}
We still consider the electric fields in their complex form: $E_{j}(t) = E_{j0}(t)e^{-2i\pi\nu_0t}$, with $E_{j0}(t)$ its complex amplitude. In the steady-state operation, the complex amplitudes of the electric fields don't depend on time, only the propagation time-delay matters which yields:
\begin{align}
E_{c0} &= \sqrt{T_1}E_{i0} - \sqrt{R_2R_1} E_{c0} \, e^{i\frac{4\pi\nu_0L}{c}}\label{eq:Ec0}, \\
E_{t0} &= \sqrt{T_2}E_{c0} \, e^{i\frac{2\pi\nu_0L}{c}} \label{eq:Et0}, \\
E_{r0} &= -\sqrt{R_1}E_{i0} - \sqrt{R_2T_1}E_{c0}e^{i\frac{4\pi\nu_0L}{c}} . \label{eq:Er0}
\end{align}

\subsection{Intracavity power}\label{sec:pcirc}

\begin{figure}[h]
		\centering
		\includegraphics[width=0.8\textwidth]{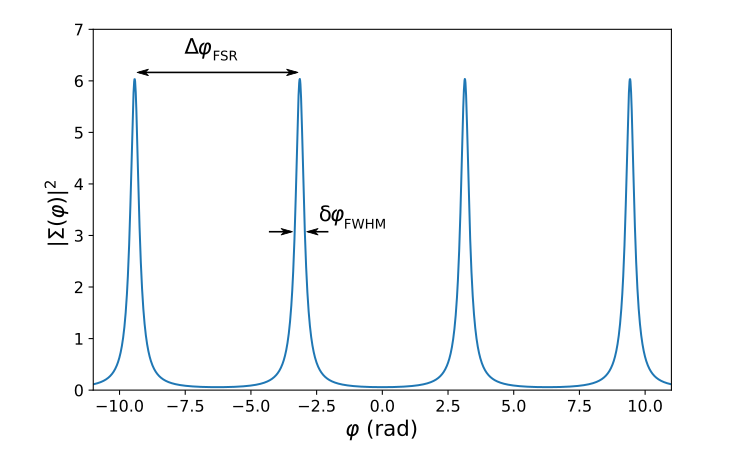}
		\caption{Squared modulus of the enhancement function as function of the round-trip phase $\varphi$ for $\sqrt{R_1} = 0.9$ and $\sqrt{R_2} = 0.914$. The spacing between the resonant peaks is noted as $\Delta\varphi_\text{FSR}$ and their full width at half maximum (FWHM) as $\delta\varphi_\text{FWHM}$.}
		\label{fig:res}
	\end{figure}

Let us now focus on the circulating field and study the intracavity power. By solving Equation \ref{eq:Ec0} for $E_{c0}$, the intracavity field writes as:
\begin{equation}
    E_{c0} = \frac{\sqrt{T_1}}{1+\sqrt{R_2R_1}e^{i\varphi}}E_{i0} = \Sigma(\varphi)E_{i0}, \label{eq:Eic}
\end{equation}
where $\varphi$ is the round-trip propagation phase:
\begin{equation}
\varphi = \frac{4\pi\nu_0L}{c} ,
\label{eq:phi}
\end{equation}
and $\Sigma(\varphi)$ is called the enhancement factor. The intracavity power is then defined as:
\begin{equation}
P_c = K|E_{c0}|^2 =K |\Sigma(\varphi)|^2 |E_{i0}|^2, 
\end{equation}
where $K=S/c\mu_0$, with $\mu_0$ the vacuum permeability and $S$ the beam cross section. 
Therefore the squared modulus of the enhancement factor determines the intracavity power according to the round-trip phase. A conventional way to write it is:
\begin{equation}
|\Sigma(\varphi)|^2 = \frac{\Sigma^2_{\text{max}}}{1 + \frac{4\mathcal{F}^2}{\pi^2}\cos^2\left(\frac{\varphi}{2}\right)} .
\label{eq:gain}
\end{equation}
This function is periodic and it reaches a maximum value $\Sigma^2_{\text{max}}$ when the cosine squared function in the denominator is equal to zero i.e. for $\varphi = \pi + 2q\pi$ with $q$ an integer (see Figure \ref{fig:res}). That is called a resonance condition. Depending on the application, it is also useful to write the resonant condition in terms of the laser frequency or the cavity length by replacing $\varphi$ by its expression (Equation \ref{eq:phi}) and by isolating either $\nu_{\textrm{res}}$ or $L_{\textrm{res}}$:
\begin{align}
\nu_{\textrm{res}} &= \frac{c}{4L} + q \frac{c}{2L} \label{eq:fsr} ,\\
L_{\textrm{res}} &= \frac{\lambda_0}{4} + q \frac{\lambda_0}{2}\label{eq:lfsr} .
\end{align}
We can then introduce the free spectral range (FSR) which represents the difference between two successive resonances. Depending on the chosen variable, it takes one of the following values:
\[\Delta\varphi_\text{FSR} = 2\pi;\ \Delta\nu_\text{FSR} = \frac{c}{2L};\ \Delta L_\text{FSR} = \frac{\lambda_0}{2}. \] 
The peak value, that is the cavity power gain, is given by:
\begin{equation}
\Sigma_{\text{max}}^2 = \frac{T_1}{(1-\sqrt{R_2R_1})^2} .
\label{eq:gmax}
\end{equation}
The other key parameter of Fabry-Perot cavities is called \textsl{finesse} and it is noted as $\mathcal{F}$ in the denominator of Equation \ref{eq:gain}. Its expression as function of the mirrors reflectivity is:
\begin{equation}
\mathcal{F} = \frac{\pi \left(R_1R_2\right)^{1/4}}{1-\sqrt{R_2R_1}} . 
\label{eq:finesse}
\end{equation}
The finesse is also defined as the ratio of the periodicity to the full width at half maximum FWHM of $|\Sigma(\varphi)|^2$ (Figure \ref{fig:res}). 
For $\sqrt{R_1}$ and $\sqrt{R_2}$ close to 1, one can see in Equation \ref{eq:finesse} that $\mathcal{F}\gg 1$. Its typical range in real cavities is from 10 to a maximum of $10^6$ and is, in practice, limited by intracavity losses which are not considered here. The higher the finesse, the sharper are the resonance peaks whose FWHM are given by:
\begin{equation}
\delta\varphi_\text{FWHM} = \frac{2\pi}{\mathcal{F}};\ \delta\nu_\text{FWHM} = \frac{\Delta\nu_\text{FSR}}{\mathcal{F}};\ \delta L_\text{FWHM} = \frac{\lambda_0}{2\mathcal{F}} .
\label{eq:fwhm}
\end{equation}
The peak height depends also on the finesse. We can re-write Equation \ref{eq:gain} for $\Sigma_{\text{max}}$ for the two Fabry-Perot cavity configurations that are used in GW detectors:
\begin{itemize}
    \item[a)] $\sqrt{R_2}=1$ and $\sqrt{R_1}\lesssim 1$ \quad \text{almost all light is reflected by the cavity},
    \item[b)] $\sqrt{R_1}=\sqrt{R_2} \lesssim 1$ \quad \text{almost all light is transmitted by the cavity}.
\end{itemize}
In the first case the end mirror is considered perfectly reflective and the input mirror reflectivity close to 1. Then, almost all the light is reflected by the cavity, the finesse becomes $\mathcal{F}\simeq \frac{\pi}{1-\sqrt{R_1}}$ and $\Sigma_{\text{max}}\simeq \frac{2}{1-\sqrt{R_1}} \simeq \frac{2\mathcal{F}}{\pi}$. In the second case $\mathcal{F}\simeq \frac{\pi}{T_1}$ and $\Sigma_{\text{max}}\simeq \frac{1}{T_1}\simeq \frac{\mathcal{F}}{\pi}$. At resonance, the light is entirely transmitted by the cavity and the reflection is zero. The arm cavities of GW detectors are of type a), and their advantage will be discussed in Section \ref{sec:cavityreflectivity}. Cavities of type b) are used to clean the spatial profile of the laser, as it will be shown in Section \ref{sec:trans}. 

\subsection{Gaussian beams} 
\begin{figure}[h]
		\centering
		\includegraphics[width=\textwidth]{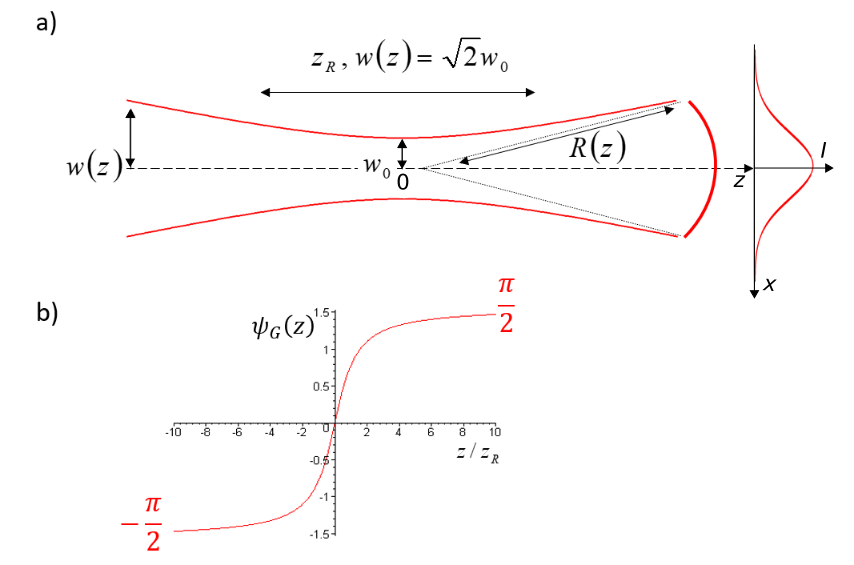}
		\caption{a) Beam profile of a gaussian beam along the propagation axis $z$ for $z_0 = 0$ (left) and at a given $z$ along the $x$ axis (right). The Rayleigh range $z_R$ represents the distance between the two points where the beam radius is $\sqrt{2}w_0$. b) Plot of the Gouy phase as function of $z/z_R$.}
		\label{fig:gb}
\end{figure}

So far, we have considered plane waves, which is a useful theoretical approximation. Now we will introduce a Gaussian beam description, which describes more accurately the spatial and phase properties of real laser beams. For that, consider an electric field propagating along the $z$ axis (the optical axis) with a complex transverse amplitude that depends on the $z$ coordinate and on the distance $r = x^2 + y^2$. For small distances and small angles with respect to the optical axis, such a wave obeys the paraxial Helmoltz Equation \cite{Siegman1986}. A Gaussian beam is the fundamental solution of that equation. One can show that the electric field complex amplitude is given by:
\begin{equation}
E(x,y,z) = E_0\frac{w_0}{w(z)} \cdot e^{i\left(\frac{2\pi}{\lambda_0}(z-z_0)+\psi_\text{G}(z)\right)} \cdot e^{-i\frac{2\pi}{\lambda_0}\frac{x^2+y^2}{2R(z)}} \cdot e^{-\frac{x^2+y^2}{w(z)^2}} \; .
\label{eq:TEM00}
\end{equation}
As one can see from the last exponential term, the transverse profile is a gaussian function whose radius $w$ depends on $z$. The gaussian peak amplitude is $E_0 w_0/w(z)$ where $w_0$ is the waist of the gaussian beam, i.e., its minimum radius obtained at the position $z=z_0$. The presence of the factor $w_0/w(z)$ in the amplitude can be understood by considering that the beam power must be the same for any $z$ so the integral on the $x-y$ plane of $\lvert E \rvert^2$ shall not depend on $z$. The first exponential term in Equation \ref{eq:TEM00} contains the propagation phase and an additional phase called Gouy phase $\psi_\text{G}(z)$ whereas the second exponential term describes the spherical phase front of radius $R(z)$. The quantities $w(z)$, $\psi_\text{G}(z)$, $R(z)$ are defined as:
\begin{align}
w(z) &= w_0\sqrt{1 + \frac{\left(z-z_0\right)^2}{z_R^2}}, \\
\psi_\text{G}(z) &= \tan^{-1}\left(\frac{z-z_0}{z_R}\right),  \\  
R(z) &= z-z_0 + \frac{z_R^2}{z-z_0},
\end{align} where $z_R=\pi w_0^2/\lambda_0$ is a parameter called Rayleigh range characterizing the beam divergence. Hence a Gaussian beam is entirely defined by the three parameters $z_0$, $w_0$ and $\lambda_0$.
Figure \ref{fig:gb} a) shows a schematic of the gaussian beam evolution while propagating along the $z$ axis. Figure \ref{fig:gb} b) shows a plot of the Gouy phase as function of the distance from the waist normalized by the Rayleigh range.
\\
A gaussian beam is resonant in a Fabry-Perot cavity when the radius of curvature of the beam matches the radius of curvature of the mirrors at their position and when the phase resonance condition is satisfied. By considering the phase term in Equation \ref{eq:TEM00}, the previously defined resonance condition on the $z$-axis writes as:
\begin{equation}
\centering
\varphi + 2\Delta\psi_G = \pi + 2q\pi,
\label{eq:res00}
\end{equation}
where $2\Delta\psi_G$ is the round-trip variation of the Gouy phase.

\begin{figure}[h!]
		\centering
		\includegraphics[width=0.7\textwidth]{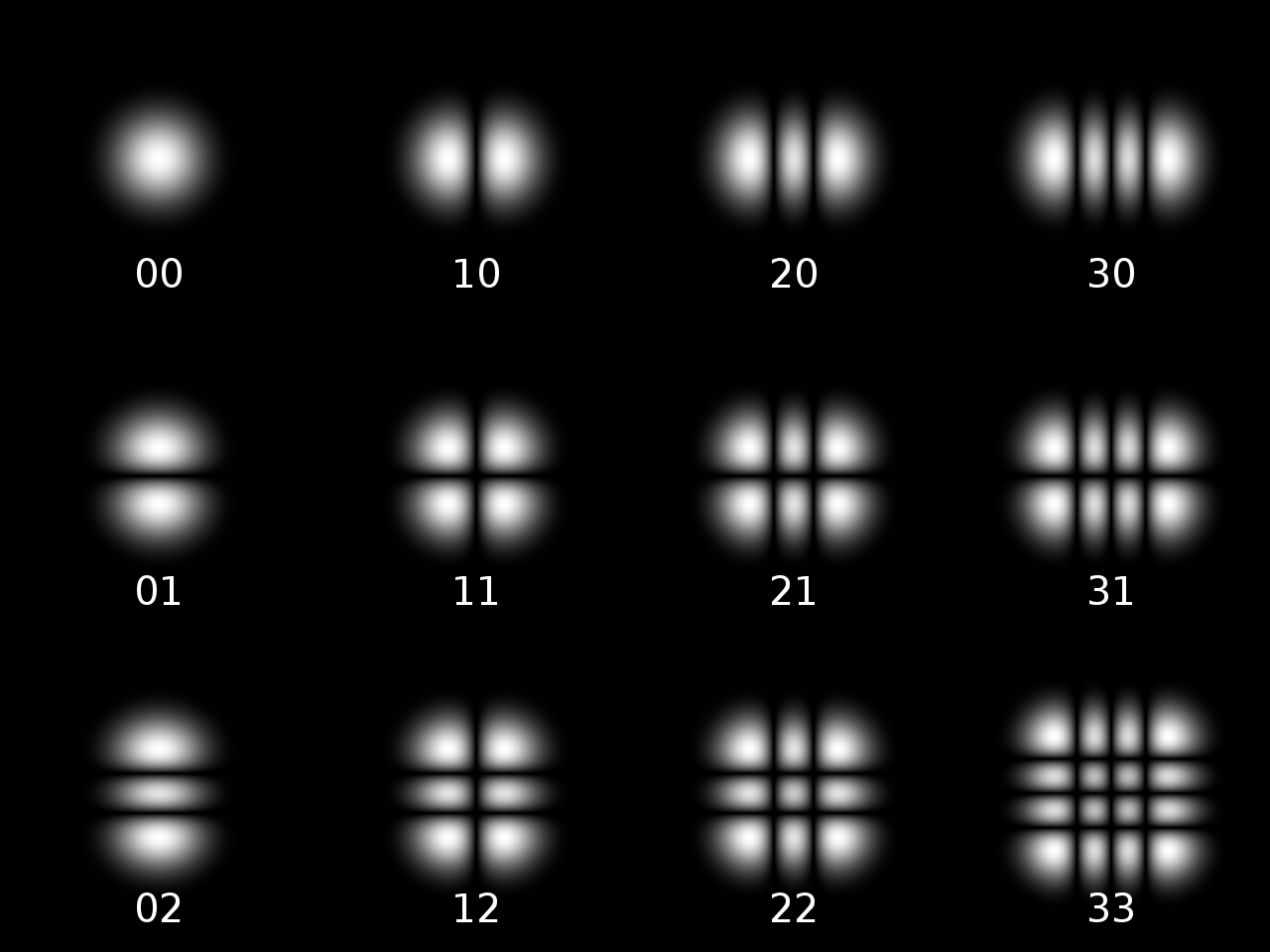}
		\caption{Intensity profiles of twelve consecutive Hermite-Gaussian modes. Each mode is defined by the integer numbers $m$ and $n$ that correspond to the number of zeros along the $x$ and $y$ axis. From \cite{wiki}.}
		\label{fig:HG}
	\end{figure}

In reality the gaussian beam in Equation \ref{eq:TEM00} is just one solution of the paraxial Helmotz equation and it represents the fundamental mode of the complete orthogonal set of eigenfunctions. One useful set are the Hermite-Gauss (HG) modes:
\begin{eqnarray}
\centering
\nonumber E_{n,m}(x,y,z) &=& \frac{E_{0}w_0}{w(z)} H_n\left(\frac{\sqrt{2}x}{w(z)}\right)H_m\left(\frac{\sqrt{2}y}{w(z)}\right)\\ &\times&e^{i\left(\frac{2\pi}{\lambda_0}(z-z_0)+(n+m+1)\psi_\text{G}(z)\right)} \cdot e^{-i\frac{2\pi}{\lambda_0}\frac{x^2+y^2}{2R(z)}} \cdot e^{-\frac{x^2+y^2}{w^2(z)}},
\label{eq:TEMnm}
\end{eqnarray}
where $H_{j}$ are Hermite polynomials of order $j$. The intensity profiles of some HG modes are shown in Figure \ref{fig:HG}, the fundamental gaussian mode corresponds to $m=n=0$.
	
HG modes can also resonate in a Farby-Perot cavity with the following resonance condition:
\begin{equation}
\label{Eq:resonance_condition}
\varphi + 2(n + m + 1)\Delta\psi_G = \pi + 2q\pi.
\end{equation}
The resonance peak of an HG mode of order ($m+n$) is therefore shifted compared to the fundamental mode (Equation \ref{eq:res00}) and this shift depends on the mode order.

\subsection{Cavity transmission and mode cleaner cavities}\label{sec:trans}

The transmitted field by a cavity is obtained by substituting Equation \ref{eq:Eic} and Equation \ref{eq:phi} in Equation \ref{eq:Et0}:
\begin{equation}
E_{t0} = \sqrt{T_2}e^{i\varphi/2} \Sigma(\varphi) E_{i0} . 
\label{eq:Et}
\end{equation}
The cavity transmission coefficient $T_C$ is defined by the ratio between the transmitted power and the input power:
\begin{equation}
T_C(\varphi) = \left| \frac{E_{t0}}{E_{i0}} \right|^2 = \frac{T_{C,\max}}{1 + \frac{4\mathcal{F}^2}{\pi^2} \cos^2\left(\frac{\varphi}{2}\right)} .\label{eq:T}
\end{equation}
Similarly to the intracavity power, the transmitted power as function of the round-trip phase $\varphi$ is periodic, exhibiting a peak when the resonant condition is satisfied. The maximum transmitted power is given by $T_\text{C,max} \times P_i$ with $P_i$ the laser input power.
If $\sqrt{R_2} = \sqrt{R_1}$ and $\sqrt{T_2} = \sqrt{T_1}$, i.e. configuration b) in Section \ref{sec:pcirc}, then $T_\text{C,max} = T_2\Sigma^2_{\text{max}} = 1$. 
That means that the cavity is transparent at resonance.

When one considers a realistic input beam with a given spatial profile, its electric field $E_i$ can be written as a linear superposition of the cavity HG eigenmodes $E_{m,n}$: 
\begin{equation}
E_i = \sum_{m,n} c_{m,n,i} \times E_{m,n} .
\end{equation}
By scanning the laser frequency or the cavity length one observes a serie of resonances, each one corresponding to a specific spatial mode (see Equation \ref{Eq:resonance_condition}). This allows a direct measurement of the modal content of the incident beam and if one could keep the cavity in a particular resonant condition with the laser, only one mode would be transmitted. For example, with a resonance on the fundamental mode, the transmitted power is $|E_t|^2_{0,0} \propto \left| c_{0,0,i} \right|^2 $ while the transmitted power of higher orders components is close to zero: $|E_t|^2_{\left(m,n\right) \neq (0,0)} \propto \frac{\left| c_{m,n,i} \right|^2}{\mathcal{F}^2} \simeq 0$
%
. In this case the cavity acts as a spatial mode-cleaner, which is used in GWDs to ensure a close to pure, stable fundamental Gaussian mode in order to maximize contrast, and minimize noise coupling from higher order modes to the output power.  
A common technique to keep the resonant condition is discussed in Section \ref{sec:PDH}.
%

%
\subsection{Cavity reflectivity}
\label{sec:cavityreflectivity}
\begin{figure}[h]
		\centering
		\includegraphics[width=\textwidth]{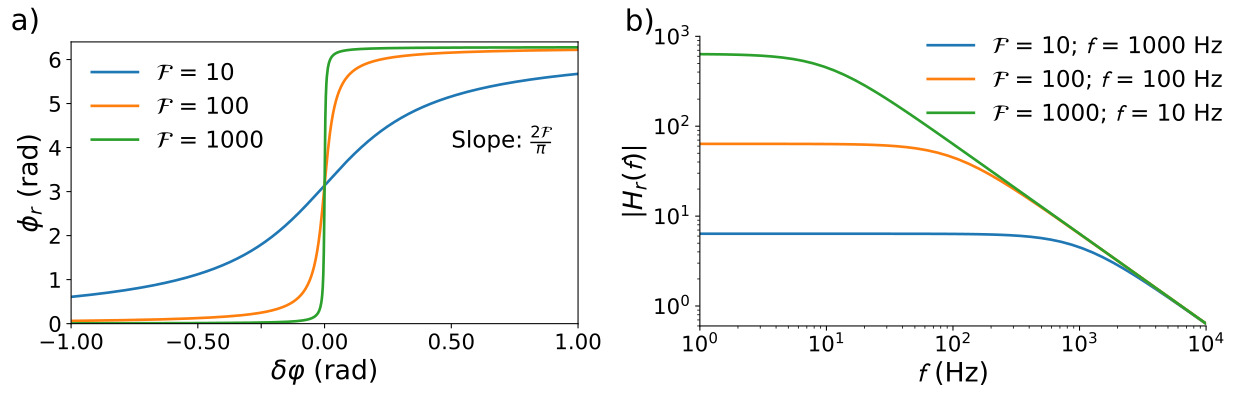}
		\caption{a) Phase of the reflected field $\phi_r$ as function of the phase deviation from resonance $\delta\varphi$ for three values of finesse and for $f\simeq 0$. b) Transfer function $H_r\left(f\right)$ for the same three finesse values. The pole frequencies correspond to an arbitrary cavity length of 7.5 km.}
		\label{fig:FPrefl}
	\end{figure}

From energy conservation, the laser power that is not transmitted by the Fabry-Perot cavity is reflected (in the absence of loss). The complex reflection coefficient of the cavity is given by $R(\varphi)=-E_{r0}/E_{i0}$ (the (-) sign is due to the convetion of Figure \ref{fig:FP}) where $E_{r0}$ can be retrieved from equations \ref{eq:Er0}-\ref{eq:Eic}. For the specific configuration where $\sqrt{R_2} = 1$ and $\sqrt{T_2} = 0$, i.e. configuration a) mentioned in Section \ref{sec:pcirc}, the transmission is zero (according to Equations \ref{eq:Et}-\ref{eq:T}) and the reflectivity $\left|R(\varphi)\right|^2$ is equal to 1. Therefore the incident beam is fully reflected for any value of the round-trip phase.\\
The important modifications happen at the phase of the reflected field around resonance. For small and slow deviations $\delta\varphi$ (slow corresponds to Fourier frequencies $f\simeq 0$) of the round-trip phase around resonance, the phase of the reflected wave with respect to the input wave is given by:
\begin{equation}
\phi_{r} = \mathrm{Arg}(R(\varphi)) \simeq \pi + 2\tan^{-1}\left(\frac{\mathcal{F}}{\pi}\delta\varphi\right) .
\end{equation}
In Figure \ref{fig:FPrefl}a) one can see the plot of the reflected phase for different values of finesse. The slope around $\delta\varphi = 0$ is equal to $2\mathcal{F}/\pi$. This is the main result that explains the enhancement of sensitivity in a Michelson interferometer with resonant Fabry-Perot cavities in its arms. Indeed, a small phase variation due to a change of the arm length is amplified by the factor $2\mathcal{F}/\pi$. 
\\ For $f\neq 0$, the Fabry-Perot  behaves as a first-order low-pass filter with a pole frequency $f_p = \frac{\delta \nu_{\text{FWHM}}}{2}$ where $\delta \nu_{\text{FWHM}}$ is given in Equation \ref{eq:fwhm}. The corresponding transfer function is:
\begin{equation}
\mathcal{H}_{r,\varphi}(f)=\frac{\phi_r}{\varphi} \simeq \frac{2\mathcal{F}/\pi}{1+i\frac{f}{f_p}}.
\end{equation}
Its module, i.e. the gain, is shown in Figure \ref{fig:FPrefl}b) for different values of the finesse. One can see that the higher the finesse is, the higher is the low frequency gain but lower is the amplification bandwidth. A trade-off between gain and bandwidth has to be done in defining the cavity parameters according to the detection goals.  
On the other hand, a small phase variation $\phi_L$ introduced by the injected laser will be reflected following the transfer function $\mathcal{H}_{r,L}\left(f\right)$:
\begin{equation}
    \mathcal{H}_{r,L}\left(f\right)=\frac{\phi_r}{\phi_L}=-\frac{1-i\frac{f}{f_p}}{1+i\frac{f}{f_p}} . 
\end{equation}
which describes a phase-shift across the cavity pole.	
\subsection{Cavity as a frequency reference: feedback loop}
\label{sec:PDH}
\begin{figure}[h]
		\centering
		\includegraphics[width=1\textwidth]{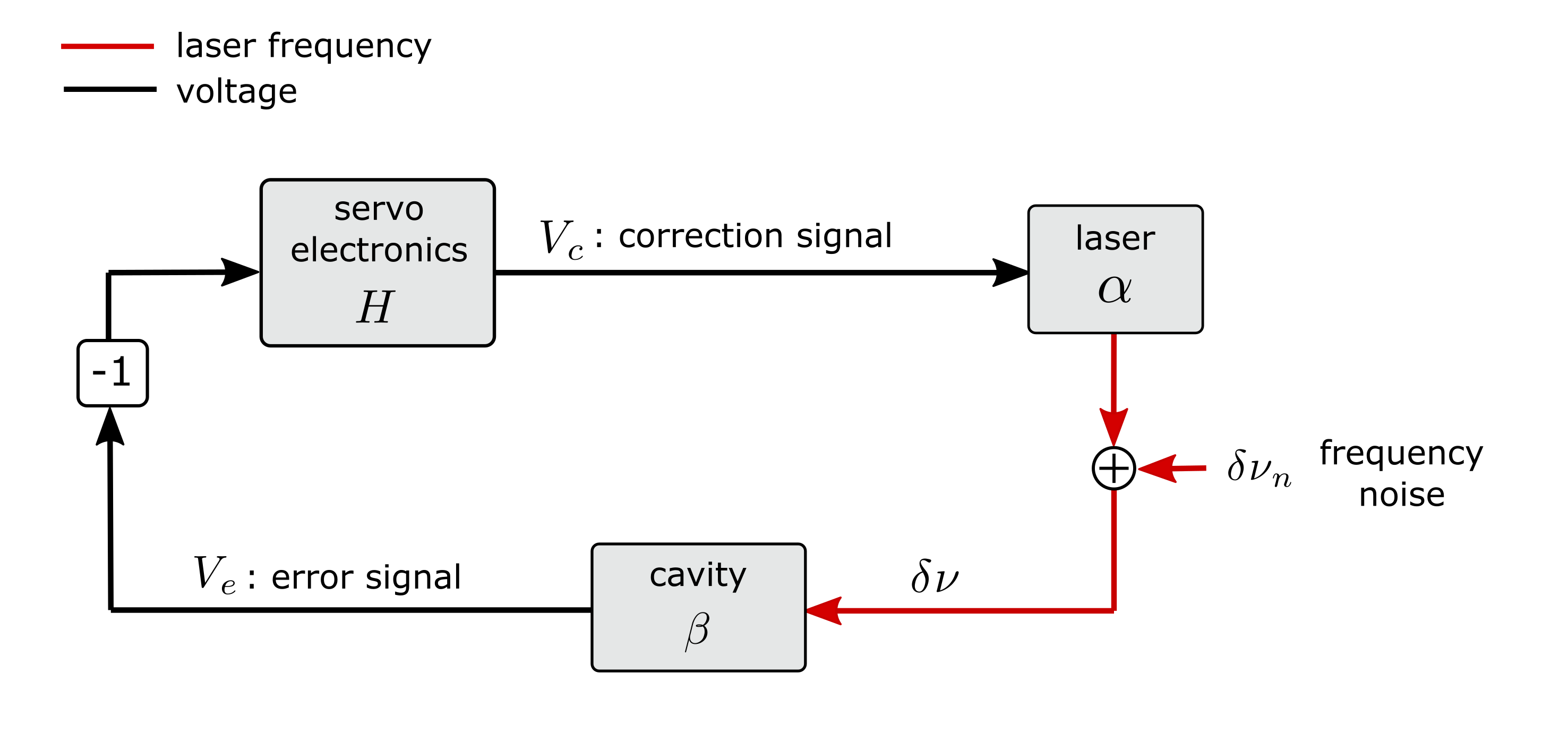}
		\caption{Block diagram describing a feedback control loop to stabilize the laser frequency to a cavity resonance.}
		\label{fig:servo}
\end{figure}

\begin{figure}[h]
		\centering
		\includegraphics[width=0.8\textwidth]{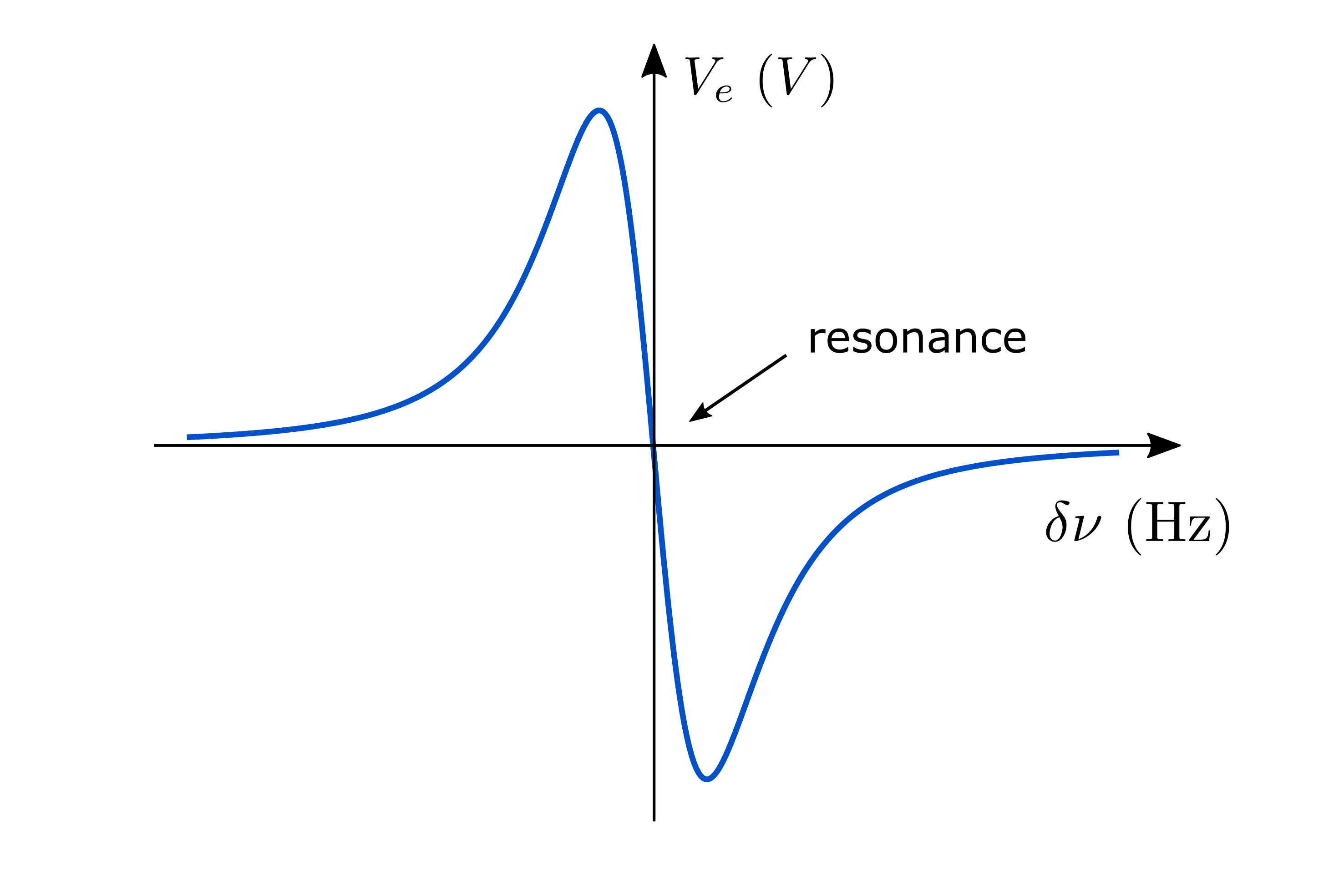}
		\caption{Error signal obtained by the PDH technique around the cavity resonance frequency. The PDH  scheme generates an error signal (in Volts) which is zero at resonance, i.e., when $\delta\nu=0$.}
		\label{fig:PDHerr}
	\end{figure}

We will now describe how to lock the laser frequency, which is fluctuating, on a cavity resonance using a feedback control loop. For this we assume a cavity in which its length is extremely stable so that the cavity resonance can be used as a frequency reference, in order to determine the laser frequency noise. \\
Figure \ref{fig:servo} shows a block diagram describing the feedback control loop. First, a laser emits on an optical frequency $\nu_0$ to which the frequency noise $\delta \nu_n$ with respect to the cavity resonance is added. At the cavity input, the total laser frequency fluctuations $\delta \nu$ are compared with the cavity resonance, and their difference will result in what is called error signal $V_e$. An electronic servo amplifies and filters the error signal to produce the correction signal $V_c$ which is sent to the laser frequency actuator.
 Let us call $\alpha$ the response of the laser frequency actuator, $\beta$ the response of the error readout scheme including the cavity, and let us assume that these responses are linear. The error signal $V_e$ is inverted to create a negative back action. It is then sent to the electronic servo which has a transfer function $H$ through which it generates the correction signal $V_c$. The parameters $V_c$, $\delta \nu$ and $\epsilon$ are linked by the following coupled equations:  	
\begin{align}
V_c &= - H \times V_e, \\
\delta \nu &= \alpha \times V_c + \delta \nu_n, \\
V_e &= \beta \times \delta \nu = \beta \times \delta \nu_n - \alpha\beta H \times V_e .
\end{align}Solved, they give:
\begin{equation}
V_e = \frac{\beta \times \delta \nu_n}{1+\alpha\beta H} . \\
\label{eq:error_signal}
\end{equation}
$H$ is designed such that, in the bandwidth of stabilization, $\alpha\beta H \gg 1$. Also, when $\left|\alpha\beta H \right|\simeq 1$, one needs $\arg\left(\alpha\beta H \right)$ far from $180^{\circ}$ in order to avoid any loop oscillation, i.e. non zero denominator in Equation \ref{eq:error_signal}. For high loop gain, $V_e \simeq \frac{\delta \nu_n}{\alpha \beta H} \to 0$ which shows that the laser remains resonant with cavity. The resonance condition \ref{eq:fsr} being fulfilled, one can write:
\begin{equation}
\label{delta_nu_delta_L}
    \left|\frac{\delta \nu}{\nu_0}\right|=\left|\frac{\delta L}{L}\right| , 
\end{equation} in the lock frequency bandwidth. One difficulty though is to generate a linear error signal. A well known adequate technique is the Pound Drever Hall (PDH)\cite{Drever1983}.
\begin{figure}[h!]
		\centering
		\includegraphics[width=\textwidth]{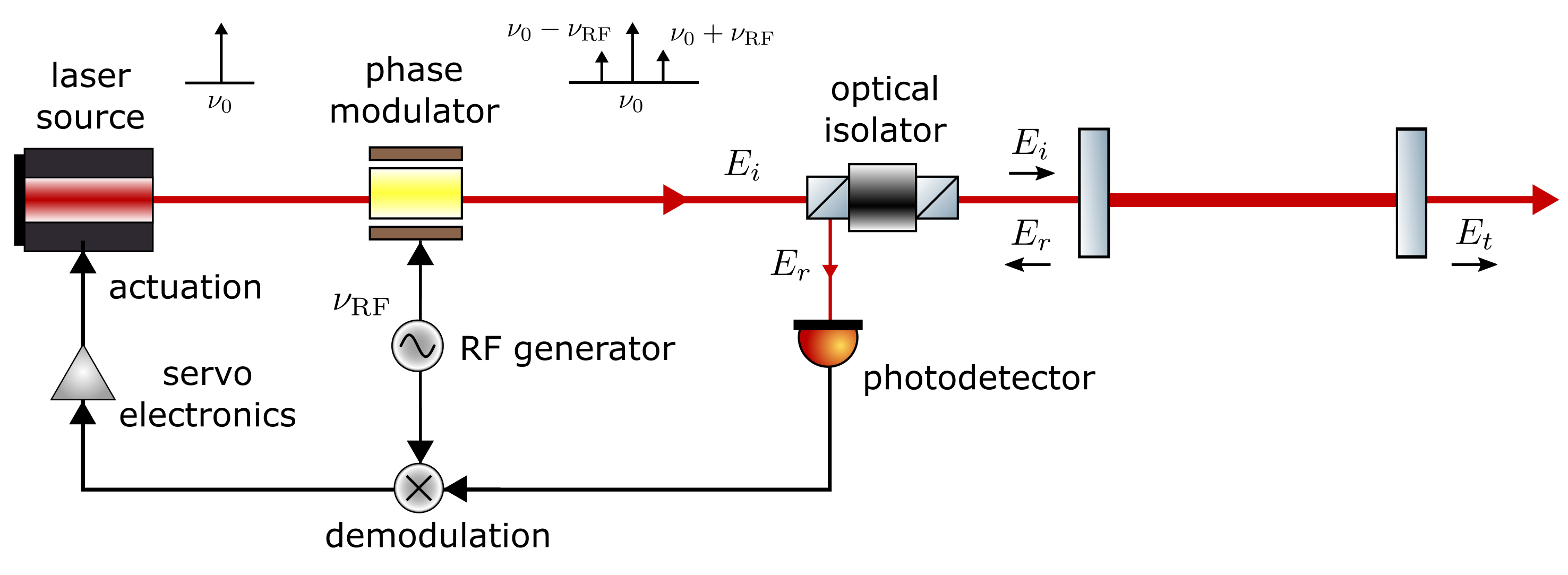}
		\caption{Schematic of the experimental implementation of the PDH technique to lock the laser frequency to a cavity resonance. On the top of the figure one can see the sideband picture representation of the optical carrier frequency of the laser $\nu_0$ and the RF sidebands $\nu_0 \pm \nu_\text{RF}$. More details in the text. }
		\label{fig:PDH}
	\end{figure}
A typical setup of this technique is depicted in Figure \ref{fig:PDH}. Before entering the cavity, the laser beam passes through a phase modulator driven by an RF (radio frequency) generator. For small modulation amplitudes only two sidebands are generated. If they are sufficiently apart from the carrier frequency $\nu_0$, the corresponding reflected field is not phase shifted when $\nu_0$ is close to the cavity resonance whereas the carrier field undergoes an important phase change (see Figure \ref{fig:FPrefl}a)). A photodetector in reflection of the cavity is then used to detect the beat note between the carrier leaving the cavity and the RF sidebands being directly reflected (not phase shifted). When the photodetector signal is demodulated at the RF modulation frequency, a linear error signal around the resonance is generated which is zero exactly at resonance, and has different signs depending if the laser frequency is above or below resonance. This error signal is plotted in Figure \ref{fig:PDHerr}. 
	
\section{Power recycled interferometer}
\begin{figure}[h]
		\centering
		\includegraphics[width=1\textwidth]{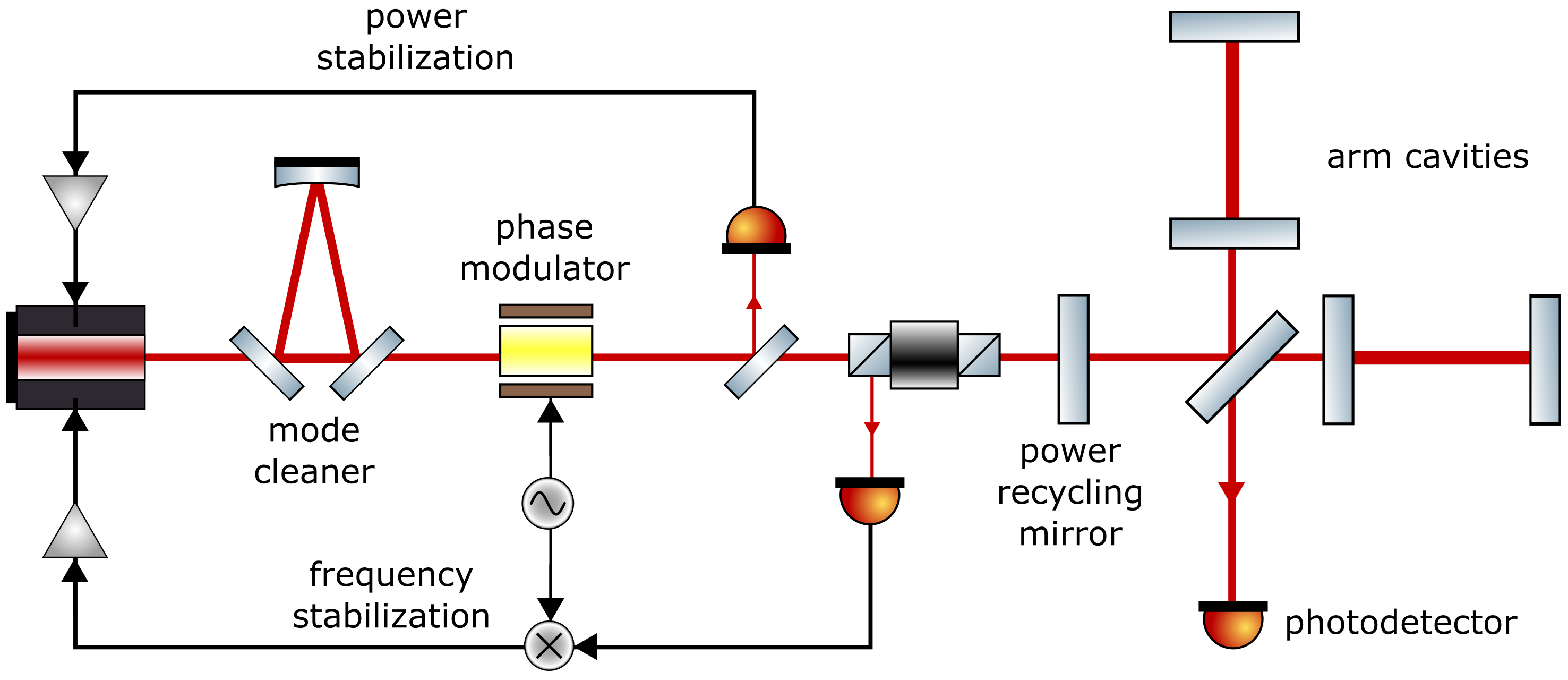}
		\caption{Schematics of a simplified gravitational wave detector, composed of a Michelson interferometer with cavities in its arms and a power recycling mirror. The figure also shows the scheme of a power stabilization in which a photodetector senses the fluctuations of a pick-off beam, the scheme of frequency stabilization in which the frequency fluctuations are read in reflection of the interferometer, and finally a triangular mode cleaner cavity.}
		\label{fig:power_recycling_detector}
	\end{figure}

Now that a simple Michelson interferometer and optical cavities were introduced, we can consider an almost complete layout of a gravitational wave detector, with its most important characteristics. This layout is shown on Figure \ref{fig:power_recycling_detector} in which we have included:
\begin{itemize}
    \item a power recycling mirror which reflects the light from the bright port (interferometer reflection port) back to the interferometer, effectively increasing the power sent to the Michelson interferometer by a gain of $G\simeq 50$ (and consequently also increasing the transmitted power by a factor of $G$), and
    \item a Fabry-Perot cavity in each arm.
\end{itemize}
With those modifications, the PSD of the signal to noise ratio expression in Equation \ref{SNR} will be:
\begin{align}
\rho^2(f) = 
\frac{
    \overline{L}^2 \cdot | \mathcal{H}_{r,\varphi}|^2 \cdot S_h(f)
}{
    \frac{4\mathcal{F}^2}{\pi^2}\Delta L_\text{DC}^2\left( S_{\text{rpn}}(f) + 
        4| \mathcal{H}_{r,L} |^2\frac{ \cdot S_\nu(f)}{\nu_0^2} +
    \frac{ h_p \nu_0}{G \overline{P}_t}\right)+ 
    4| \mathcal{H}_{r,\varphi}|^2 
        S_{\delta L_-} }  . 
        \label{SNR_detector}
\end{align}With respect to Equation \ref{SNR}, the last equation sees the contribution of the cavity responses $\mathcal{H}_{r,\varphi}(f)$ and $\mathcal{H}_{r,L}(f)$, an enhancement of the arm length offset $\Delta L_{DC}$ by a factor $2\mathcal{F}/\pi$ and the enhancement of the incident power on the interferometer by the power recycling factor $G$.

Also are included in the figure triangular cavity (input mode cleaner) with the main role to increase the purity of the beam sent to the interferometer,\footnote{The input mode cleaner also reduces the beam pointing noise which is responsible for coupling energy into high order modes.} and frequency and power stabilization control loops to reduce their corresponding noise contribution. These will be detailed in the next two sections.
\subsection{Frequency stabilization}
As previously shown, frequency noise can limit the sensitivity of GWDs. It is therefore necessary to implement a feedback control loop to stabilize the laser frequency. In these detectors, one of the main control loops  uses the interferometer's common mode as a frequency reference, as illustrated in Figure \ref{fig:power_recycling_detector}. To the first order of $\Delta L$ and according to Equation \ref{eq:refl_ITF}, only the phase of the reflected beam senses the interferometer:
\begin{equation}
    \bar{\phi}\left(t\right)\simeq \frac{4\pi\nu_0\bar{L}}{c}-\frac{4\pi\nu_0\Delta L}{c}h\left(t\right)-\phi_L\left(t-\frac{2\bar{L}}{c}\right).
\end{equation}in which we can obviousely neglect the GW effect to obtain:
\begin{equation}
    \bar{\phi}\left(t\right)\simeq \frac{4\pi\nu_0\bar{L}}{c}-\phi_L\left(t-\frac{2\bar{L}}{c}\right).
\end{equation}Therefore, the interferometer reflects the incident beam as a single arm of lenth $\bar{L}$. By placing identical arm cavities, and employing the PDH tehcnique, one senses the frequency noise $\delta\nu$ with respect to a virtual cavity of length $\bar{L}$. After stabilization, within the control loop bandwidth, the residual frequency fluctuations follows Equation \ref{delta_nu_delta_L} and one gets:
\begin{equation}
    \frac{S_{\nu}\left(f\right)}{\nu_0^2}\simeq\frac{S_{\delta L_+}\left(f\right)}{\bar{L}^2} , 
\end{equation}where $S_{\delta L_+}\left(f\right)$ is the PSD of the sum of the length fluctuations from both interferometer's arms. Since (i) $S_{\delta L_+}\left(f\right)\simeq S_{\delta L_-}\left(f\right)$(most of the length fluctuations are uncorrelated between the two arms), (ii) $\mathcal{F}\times\Delta L_\text{DC}\ll \bar{L}$ (iii) $| \mathcal{H}_{r,\varphi}|$ is of the same order of magnitude as $| \mathcal{H}_{r,L} |$ in the frequency band where $S_{\delta L_-}$ is dominant, we can neglect the frequency noise term in Equation \ref{SNR_detector} with respect to the length fluctuations. In a perfectly symmetric interferometer, frequency noise does not couple to the output port. The coupling factor due to unequal armlengths is of the order of $\mathcal{F}\times \Delta L_\text{DC}/ \bar{L}\ll\mathcal{F}\times\lambda_0/\bar{L}\simeq 3\times 10^{-7}$. In practice, other asymmetries (mainly due to the finesse difference between the two arms) are to be taken into account and the coupling factor is of the order of $10^{-2}$. This value is small enough for the stabilized frequency noise to not affect the detector sensitivity. 

\subsection{Power stabilization}
Another necessary stabilization control loop in GWDs is the power stabilization one. The usual technique for power stabilization consists in sensing the power fluctuations of a pick-off beam directly with a photodetector, and use a feedback control loop (similar to the one described in Section \ref{sec:PDH}) like shown on Figure \ref{fig:power_recycling_detector}. Let $R$ be the power ratio of the pick off, and $P$ the power of the main beam in transmission of the phase modulator. In the time domain, the power fluctuations $\delta P_d$ sensed by the power stabilization photodetector are:
\begin{equation}
    \delta P_d = R\left(\delta P_\text{tech} + \delta P_\text{corr}\right) + \delta P_{\textrm{SN}}\left(R \bar{P}\right) ,
\end{equation}where $\delta P_\text{tech}$ represents the technical power fluctuations of the main beam, $\delta P_\text{corr}$ are the power corrections injected by the stabilization loop and $\delta P_{\textrm{SN}}\left(R \bar{P}\right) $ is the shot noise corresponding to the mean power $R\bar{P}$. In the bandwidth of stabilization, $\delta P_d\simeq 0$, the residual power fluctuations of the main beam are given by:
\begin{equation}
    \delta P = \delta P_\text{tech} + \delta P_\text{corr}+\delta P_{\textrm{SN}}\left(\bar{P}\right)=-\frac{\delta P_{\textrm{SN}}\left(R \bar{P}\right)}{R}+\delta P_{\textrm{SN}}\left(\bar{P}\right) .
\end{equation}In the last equation, both shot noises are uncorrelated and since $R \ll 1$, we have in terms of PSD:
\begin{equation}
    S_{\delta P}\left(f\right)=\frac{h_p \nu_0 R \bar{P}}{R^2}+h_p \nu_0 \bar{P}\simeq \frac{h_p \nu_0 \bar{P}}{R} . 
\end{equation} Finally, the PSD of the relative power noise will be given by:
\begin{equation}
    S_{\textrm{rpn}}\left(f\right) = \frac{h_p\nu_0}{R \bar{P}}.
\end{equation} This equation shows that, in order to reduce (and possibly neglect) $S_{\textrm{rpn}}$ in Equation \ref{SNR_detector}, the pick off power $R \bar{P}$ needs to be much larger than the power transmitted by the interferometer, i. e. $R \bar{P}\gg G \bar{P}_t$. In practice, there is one to two order of magnitudes between both.\footnote{In practice, this is true once the transmitted beam is filtered by an additional cavity called Output Mode Cleaner (OMC) which transmits the matched fundamental modes between the two arms.}

\subsection{Shot noise sensitivity curve}

%
Thanks to the power and the frequency stabilization control loops, Equation \ref{SNR_detector} reduces to:
\begin{eqnarray}
\label{SNR_detector_after_stabilization}
\nonumber \rho^2(f) &=& \frac{\bar{L}^2 \cdot \left|\mathcal{H}_{r,\varphi}\left(f\right)\right|^2\cdot S_h(f)}{\left|\mathcal{H}_{r,\varphi}\left(f\right)\right|^2 \times 4 S_{\delta L_-}(f) +  \frac{4\mathcal{F}^2}{\pi^2}\Delta L_\text{DC}^2 h_p \nu_0 / \left(G\bar{P_t}\right) } . \\
\end{eqnarray}The signal-to-shot-noise ratio of the power recycled detector with arm cavities will be given by:

\begin{equation}
\rho^2_{\text{SN}}(f) =  \frac{64\mathcal{F}^2G\bar{P_0} \bar{L}^2}{h_p c \lambda_0}\left|\frac{1}{1 + i\frac{f}{f_p}}\right|^2S_h(f) .
\end{equation}
By setting the signal-to-shot-noise ratio equal to 1, i.e., $\rho^2_{\text{SN}}(f) =1$, we infer the PSD of the shot noise limit in strain sensitivity:
\begin{equation}
\label{Shot}
S_{h,\text{SN}}(f) = \frac{h_p c \lambda_0}{64\mathcal{F}^2G\bar{P_0} \bar{L}^2}\left(1 + \frac{f^2}{f_p^2}\right) .
\end{equation}Note that the sensitivity of the interferometer can be improved by increasing the interferometer power.
In the following sections we will calculate the coupling of length noise and quantum noise to the strain sensitivity of a GWD, and finally calculate the full sensitivity of the detector described in this last section (Figure \ref{fig:power_recycling_detector}).
\section{Other noise contributions}

\subsection{Length noise contributions}
As already discussed, and also shown by Equation \ref{SNR_detector_after_stabilization}, differential length noise $S_{\delta L_-}$ can limit the interferometer sensitivity. In the following, we will estimate the contributions of different sources to differential length noise by adopting a simple approach based on the harmonic oscillator model. 
\subsubsection{Harmonic oscillator}

\begin{figure}[h]
\centering
\includegraphics[width=\textwidth]{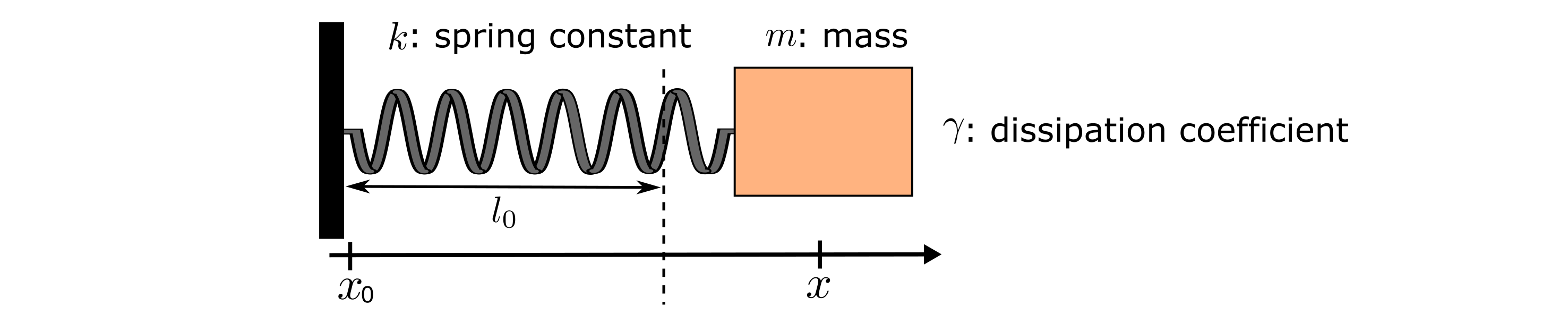}
\caption{Representation of an harmonic oscillator as a block of mass attached to a spring. $x$ is the mass position, $x_0$ is the wall position and $l_0$ is the length of the unloaded spring at rest.}
\label{fig:mass_spring}
\end{figure}

\begin{figure}[h]
\centering
\includegraphics[width=\textwidth]{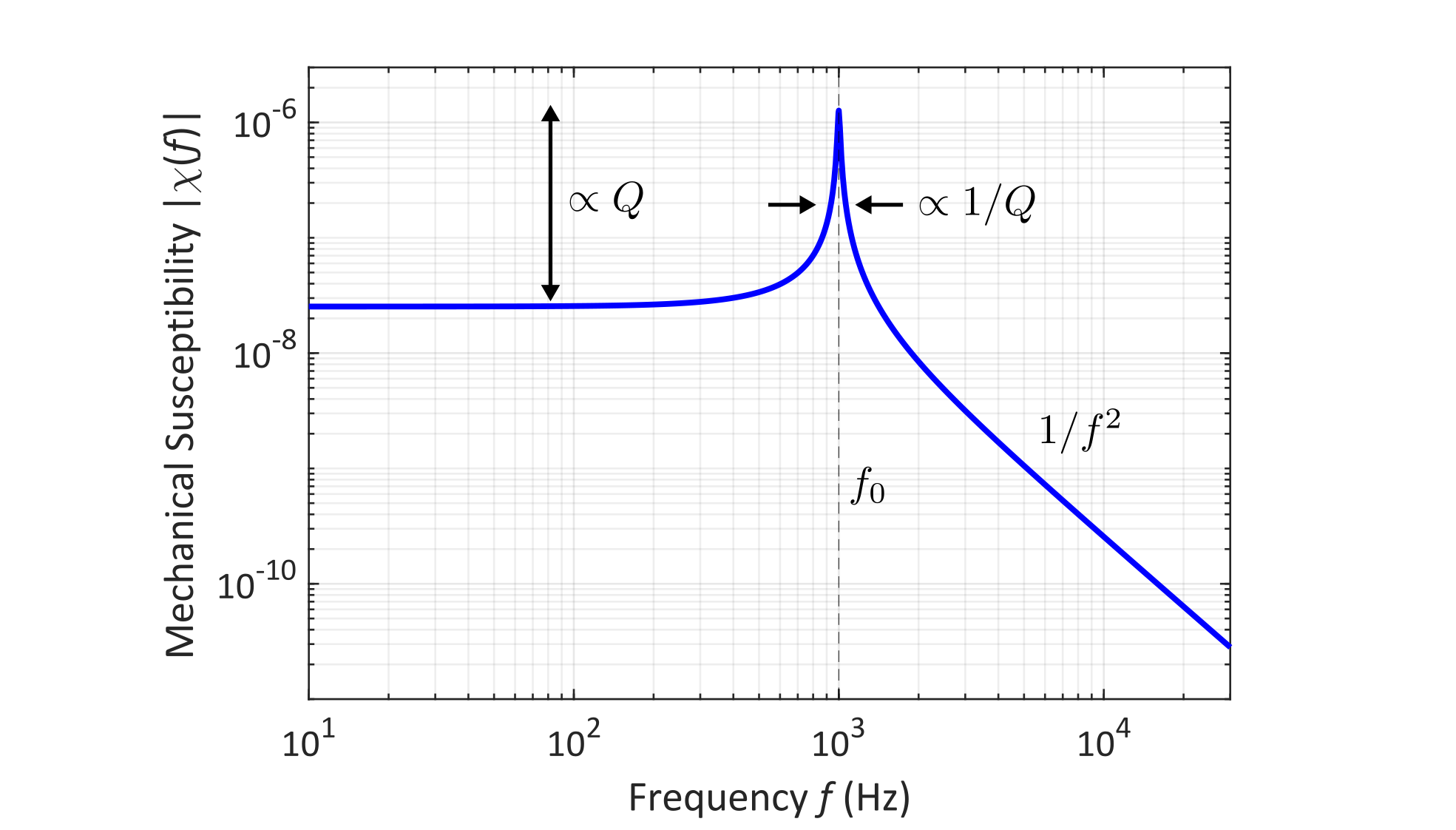}
\caption{Magnitude of the mechanical susceptibility of a harmonic oscillator with a fundamental resonance frequency $f_0$ equal to 1 kHz and a quality factor $Q$ as a function of the Fourier frequency $f$.}
\label{fig:harmonic_oscillator}
\end{figure}

Figure \ref{fig:mass_spring} illustrates a harmonic oscillator which is represented by a point like mass $m$ attached to a moving wall by a spring with spring constant $k$ and with unloaded length at rest $l_0$. This mass can be displaced along the spring axis by applying an external force $F_0$ or by changing the position of its attachment point $x_0$ which we will consider to be time dependent. 
We assume that this oscillator is damped, with a fluid damping coefficient proportional to the velocity. The dynamics of this damped harmonic oscillator subject to an external driving force $F_0$ can be described with Newton's second law in the time domain as:
\begin{equation}
\label{2ndLaw_Harmonic_Oscillator}
 m\frac{d^2x(t)}{dt^2}  = F_0 - k(x(t)-x_0(t)-l_0) - m\gamma \frac{dx(t)}{dt}    .
\end{equation}
The second term after the equality is the restoring spring force, which is proportional to the spring extension. The last term represents the damping, with $\gamma$ being the viscous damping coefficient. Assuming that the system has a linear response to a sinusoidal external force, one can obtain the equation of motion in the frequency domain via a Fourier Transform.  
The system's frequency response is characterized by the mechanical susceptibility $\chi(f)$, in units of m/N, given by:
\begin{align}
\label{mechanical_susceptibility}
\chi(f) &=  \frac{x(f)}{F_0(f) + k x_0(f)} = \frac{1/m}{\omega_0^2-\omega^2+i\frac{\omega\omega_0}{Q}} ,
\end{align}
where $\omega = 2\pi f$ is the angular frequency, $\omega_0 = 2\pi f_0 = \sqrt{k/m}$ is the fundamental angular resonance frequency, and $Q = \omega_0 / \gamma$ is the mechanical quality factor. Equation \ref{mechanical_susceptibility} shows that the mass $m$ can similarly be excited by the force $F_0$ or the motion of the fixation point $x_0$.  
Figure \ref{fig:harmonic_oscillator} shows a plot of the magnitude of the mechanical susceptibility as a function of frequency. From this plot, one can see that the system response is frequency independent for low frequencies ($f \ll f_0$), it has a peak response at resonance, and then the response decreases proportionally to $1/f^2$.\\ 
The momentum $p$ and the total mechanical energy (Hamiltonian $H$) of the harmonic oscillator are given by:
\begin{align}
p &= m\frac{dx}{dt}  ,\\[1em]
\label{energy_harmonic_oscillator}H &= \frac{p^2}{2m} + m\frac{\omega_0^2x^2}{2}   ,
\end{align}where $\left(x,p\right)$ represents the conjugate canonical variables \cite{Landau1976}. 

\begin{figure}[h]
\centering
\includegraphics[width=\textwidth]{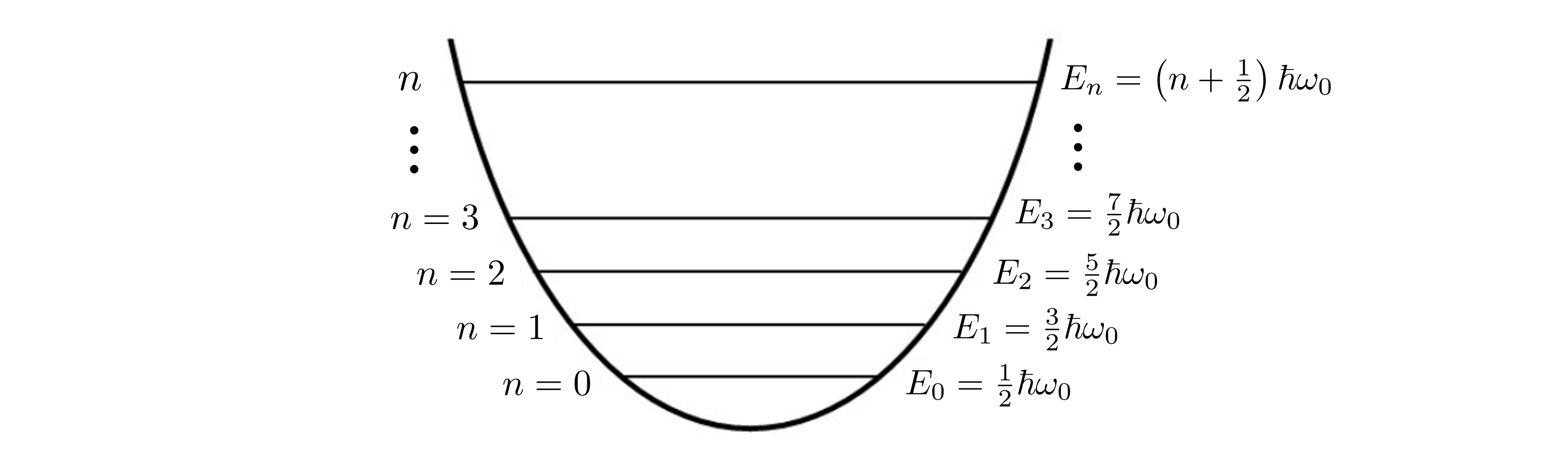}
\caption{Quantized energy states of a harmonic oscillator of resonance frequency $\omega_0$. The minimum energy state is non zero.}
\label{fig:Harmonic_oscillator_energy_states}
\end{figure}

The quantization of the harmonic oscillator can be inferred by associating to $\left(x,p\right)$ the conjugate observables $\left(\hat{x},\hat{p}\right)$ obeying the commutation relationship:

\begin{equation}
    \left[\hat{x},\hat{p}\right]=i\hbar\quad;\quad\hbar = \frac{h_p}{2\pi} , 
\end{equation}and the observable energy is given by the operator Hamiltonian:
\begin{equation}
\hat{H} = \frac{\hat{p}^2}{2m} + m\frac{\omega_0^2\hat{x}^2}{2}   . 
\end{equation}With some algebra \cite{Cohen1973}, one can show that by defining the observable:
\begin{equation}
    \hat{N}=\frac{1}{2}\left(\sqrt{\frac{ m\omega_0}{\hbar}}\hat{x}+i\frac{1}{\hbar m\omega_0}\hat{p}\right)\left(\sqrt{\frac{m\omega_0}{\hbar}}\hat{x}-i\frac{1}{\hbar m\omega_0}\hat{p}\right) ,
\end{equation}the Hamiltonian can then be written as:
\begin{equation}
    \hat{H}=\hbar\omega_0\left(\hat{N}+\frac{1}{2}\right) ,
\end{equation}and that $\hat{N}$ has integer eigenvalues $\{n\geq 0\}$. This shows that the harmonic oscillator has a set of discrete energy states given by (see Figure\ref{fig:Harmonic_oscillator_energy_states}):
\begin{equation}
    E_n=\hbar\omega_0 \left(n+\frac{1}{2}\right) .
\end{equation}A fundamental comment is to be made here: the lowest energy state has a non zero energy $E_0=\hbar\omega_0/2$ due to the Heisenberg uncertainty principle. We'll show later that this is responsible for quantum noise in the GWD. 

This summarizes all the properties of a classical and quantum harmonic oscillator that we need to set a simple description of different sources of differential length noise in GW detectors and how to mitigate them. 

\subsubsection{Seismic noise}
\begin{figure}[h]
\centering
\includegraphics[width=\textwidth]{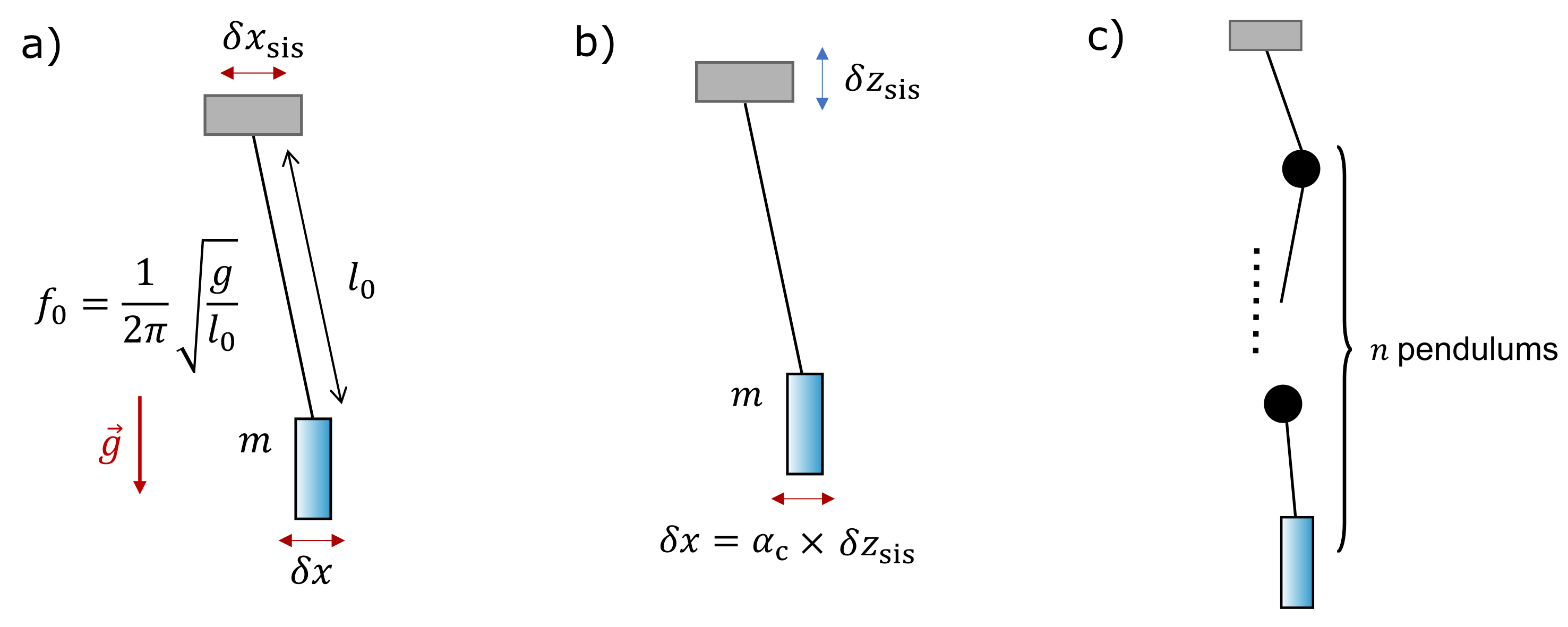}
\caption{Representation of coupling from seismic noise displacement $\delta x_\text{sis}$ at the top of a pendulum mirror to mirror displacement $\delta x$ for: a) single pendulum driven by horizontal seismic noise, b) single pendulum driven by vertical seismic noise, and c) a chain of $n$ pendulums.}
\label{fig:pendulum}
\end{figure}

%
In this section, we will analyse the seismic noise contribution to GWDs. We will refer to seismic noise as all types of vibrational noise that undergo the mirrors through their contact to the ground. The GWD mirrors are in free fall at the detection bandwidth and are suspended as pendulum mirrors. In the frequency band of interest, a pendulum mirror behaves as an harmonic oscillator under small oscillations with a spring constant $k=mg/l_0$ where $g$ stands for the local gravitational acceleration and $l_0$ its length. The pendulum acts like a passive seismic attenuator for frequencies larger than the fundamental resonance frequency (see Figure \ref{fig:harmonic_oscillator}). Let us now calculate the horizontal displacement $\delta x$ of the mirror substrate illustrated on Figure \ref{fig:pendulum} a).
For frequencies $f \geq 1 \text{ Hz}$, the ASD of the horizontal acceleration noise $\delta a(f)$ and of the seismic horizontal displacement ground noise $\delta x_\text{sis,h}$ are typically equal to:
\begin{align}
&\delta a(f) \simeq 4 \times 10^{-6} \text{ ms}^{-2}/\sqrt{\text{Hz}}, \\
&\delta x_{\text{sis}}(f) \simeq 10^{-7} \left(\frac{1\,\text{Hz}}{f}\right)^2 \text{ m}/\sqrt{\text{Hz}}. \label{eq:horizontal_seismic}
\end{align}
The suspension for a Virgo mirror has a length of 70 cm, which will lead to a fundamental resonance frequency of 0.6 Hz. For an ideal suspension, the mirror's horizontal displacement $\delta x(f)$ can be calculated by using the mechanical susceptibility (Equation \ref{mechanical_susceptibility}). This will lead to:
\begin{equation}
\delta x_\text{h}(f) = |\chi(f)| \times m\omega_0^2 \, \delta x_{\text{sis}}(f) , 
\label{eq:seismic_trans}
\end{equation}
which, for $f \gg f_0$, can be approximated to:
\begin{equation}
\delta x_\text{1pend,h}(f) \simeq \frac{f_0^2}{f^2} \delta x_{\text{sis}}(f) \simeq 10^{-7} \,\left(\frac{f_0}{f}\right)^2 \times \left(\frac{1\,\textrm{Hz}}{f}\right)^{2} \text{ m}/\sqrt{\text{Hz}} .    
\label{eq:seismic_1pendulum}
\end{equation}
Cascading two pendulum suspensions enhances the attenuation even further. The mirror displacement $ \delta x_\text{2pend}$ now can be calculated by inserting the attenuated displacement of the second mirror base (Equation \ref{eq:seismic_1pendulum}) into Equation \ref{eq:seismic_trans}. This will lead to: 
\begin{equation}
 \delta x_\text{2pend,h}(f) \simeq \left(\frac{f_0}{f}\right)^{4} \delta x_{\text{sis}}(f) \simeq 10^{-7} \,\left(\frac{f_0}{f}\right)^4 \times \left(\frac{1\,\textrm{Hz}}{f}\right)^{2} \text{ m}/\sqrt{\text{Hz}} . 
\end{equation}
This equation can be generalized for a system with $n$ pendulums (see Figure \ref{fig:pendulum}c)) as:
\begin{equation}
 \delta x_{n\text{pend,h}}(f) \simeq \left(\frac{f_0}{f}\right)^{2n} \delta x_{\text{sis}}(f) \simeq 10^{-7} \, \left(\frac{f_0}{f}\right)^{2n} \times \left(\frac{1\,\textrm{Hz}}{f}\right)^{2} \text{ m}/\sqrt{\text{Hz}}.   
\end{equation}
As shown in Figure \ref{fig:pendulum}b) seismic noise driving the top base of the mirror vertically ($\delta z_{\text{sis}}$) will also couple to horizontal mirror displacement by:
\begin{equation}
\delta x_\text{v}(f) = \alpha_{c} \times \delta z_{\text{sis}}(f) ,
\end{equation}
with a coupling coefficient of approximately $\alpha_{c} \simeq 1\%$ mainly related to mechanical asymmetries. To mitigate its effect, vertical isolation is required. As for the pendulums, vertical filters are based on passive vertical suspensions behaving like harmonic oscillators in the frequency of interest. In the case of the Virgo detector, the number of horizontal filters pendulums is $n=7$ and the vertical filters are $n=5$. The ASD of the total horizontal mirror displacement is an uncorrelated sum between the displacement due to horizontal and vertical seismic noise. In the case of the Virgo detector is given by:  
\begin{align}
\delta x(f) = \sqrt{\left[\delta x_{\text{sis}}(f) \times \left(\frac{f_0}{f}\right)^{14}\right]^2 + \left[\delta z_{\text{sis}}(f) \times \left(\frac{f_0}{f}\right)^{10} \times \alpha_{c}^5\right]^2} . 
\end{align}
%
%

The PSD of the equivalent strain due to seismic noise can again be computed by setting the signal to (seismic) noise ratio in Equation \ref{SNR_detector} equal to 1. Considering the 4 suspended mirrors of the arm cavities of the detector, the total seismic noise projected to strain sensitivity is given by Equation \ref{SNR_detector_after_stabilization}:
\begin{equation}
\label{seismic}
S_{h,\text{sis}}\left(f\right) \simeq \frac{4\times4}{\bar{L}^2}\left[\delta x_{\text{sis}}(f) \times \left(\frac{f_0}{f}\right)^{14}\right]^2 + \left[\delta z_{\text{sis}}(f) \times \left(\frac{f_0}{f}\right)^{10} \times \alpha_{c}^5\right]^2     .
\end{equation}In practice, $\delta z_{\text{sis}}(f)$ and $\delta x_{\text{sis}}(f)$ are of the same order of magnitude.
\subsection{Thermal noise}

Another important source of differential displacement noise is thermal noise, which sets a limitation in the degree to which the suspended mirror can stay at rest with the system in equilibrium at a temperature $T$. In order to introduce the physics underneath the thermal noise, we will follow the Langevin approach of the Brownian motion \cite{Langevin1908}. 
\begin{figure}[h]
\centering
\includegraphics[width=1\textwidth]{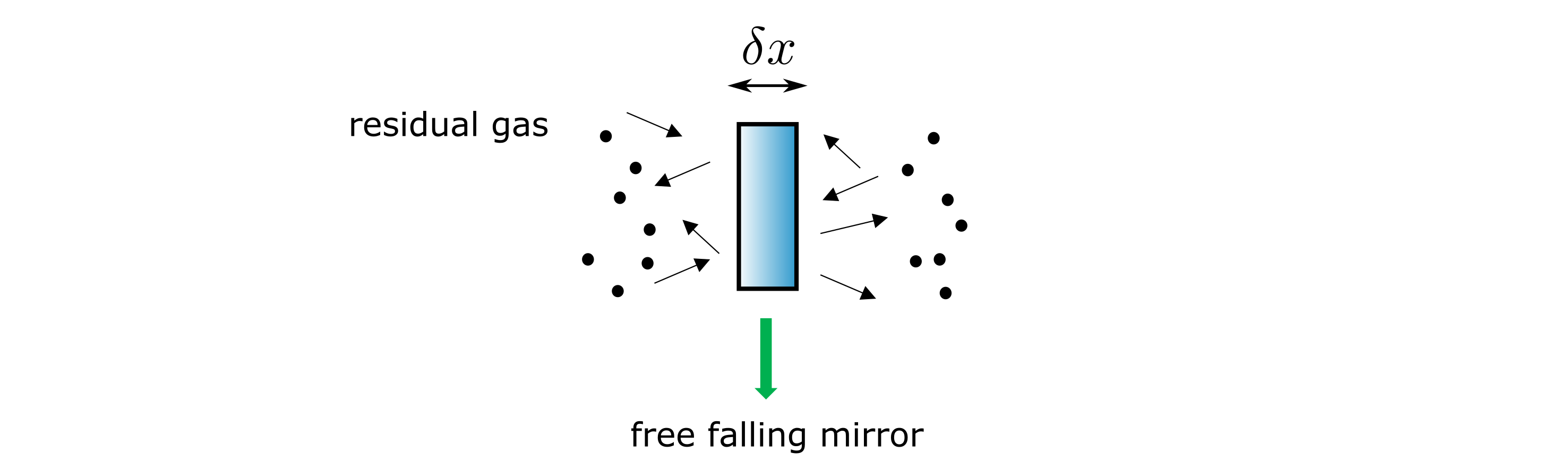}
\caption{Representation of a free falling mirror surrounded by residual gas particles which imprints a random momentum in the mirror position.}
\label{fig:thermalnoise}
\end{figure}
Let us consider one free falling mirror surrounded by a gas of particles each transferring a random momentum to the mirror, as shown in Figure \ref{fig:thermalnoise}. This momentum transfer results into a driving stochastic force $F_L\left(t\right)$ called Langevin Force. The gas is also responsible for a viscous damping proportional to the mirror velocity as described in Equation \ref{2ndLaw_Harmonic_Oscillator}. For the motion of the mirror along the $x$ axis one can simply use the damped harmonic oscillator model driven by $F_L$:
\begin{align}
m\frac{d^2x(t)}{dt^2} = m\frac{dv(t)}{dt} = -m \gamma v(t) + F_L(t) . 
\label{eq:langevin}
\end{align}
The same equation stands for the other directions.
Thermal noise is a stochastic process in which the mean value is zero at all times, i.e., $\langle F_L(t) \rangle = 0$. This reflects the fact that the mirror is equally pushed in any direction by the kicks from the gas particles. It is also a Markovian process, i.e. $\Gamma_{F_L}(\tau) = \langle F_L\left(t\right)F_L\left(t+\tau\right)=\sigma_{F_L}^2\times\Delta t\times\delta(\tau)$ where $\Delta t$ is a very short amount of time above which the correlation vanishes\footnote{the delta function can be approximated by a gate function of width $\Delta t$ and height $1/\Delta t$} (see Figure \ref{white_colored_noise}). This equation shows that the force values at different times are uncorrelated.

The damping coefficient $\gamma$ and the driving force $F_L$ have the same physical origin, but what is the relationship between them? To answer this question let us now solve Equation \ref{eq:langevin} for the velocity, that will lead to the general solution:
\begin{equation}
v(t) = v_0e^{-\gamma t} + \frac{1}{m}\int_0^t e^{-\gamma(t-t')} F_L(t')dt' , 
\end{equation}
where $t=0$ was chosen such as $F_L(t=0) = 0$ and $v(t=0) = v_0$. From this equation, one can see that the velocity mean value will be $\langle v(t) \rangle = v_0e^{-\gamma t}$ since $\langle F_L(t) \rangle = 0$\footnote{The time average is taken over times much shorter than $1/\gamma$}. In addition to that, the variance of the velocity can be calculated:
\begin{align}
\nonumber \sigma_v^2(t)& = \langle \left(\frac{1}{m}\int_0^t e^{-\gamma(t-t')}F_L(t')dt'\right)^2 \rangle  \\
\nonumber & = \frac{1}{m^2}\int_0^t dt'\int_0^{t'} dt'' \langle F_L(t')F_L(t'')\rangle e^{-\gamma(t-t')}e^{-\gamma(t-t'')} , \\
 & = \frac{\Delta t\,\sigma_{F_L}^2}{4\gamma m^2}(1-e^{-2\gamma t}) . \label{eq:variance_v}   \end{align}
The mean kinetic energy of the mirror in thermodynamic equilibrium can now be calculated, which, according to the equipartition theorem, will result in: 
\begin{equation}
\langle U_k(t) \rangle = \frac{1}{2}m\langle v(t)^2 \rangle \xrightarrow{t \to \infty} \langle U_k(t) \rangle = \frac{1}{2}m\sigma_v^2(t \to \infty) = \frac{k_BT}{2} , \label{eq:mean_Uk}   
\end{equation}
with $k_B$ being the Boltzmann's constant. Now, substituting $\sigma_v^2$ of Equation \ref{eq:variance_v} into Equation \ref{eq:mean_Uk}, will lead to: 
\begin{equation}
\sigma_{F_L}^2 = \frac{1}{\Delta t}\times 4m\gamma k_BT=\int_\frac{1}{\Delta t}4m\gamma k_BT df ,
\end{equation}The Langevin force PSD is then:
\begin{equation} 
S_{F_L}(f) = 4m\gamma k_BT  .
\end{equation}
Taking into account the mechanical susceptibility for a free falling mirror:
\begin{equation}
\chi(f) = \frac{1/m}{-\omega^2+i\omega\gamma} = \frac{-\omega -i\gamma}{m\omega(\omega^2+\gamma^2)} ,
\end{equation}
one can calculate the PSD of the Brownian motion of the mirror:
\begin{equation}
S_{x}(f) = |\chi(f)|^2S_{F_L}(f) = \frac{4k_BT}{\omega}\left|\text{Im}(\chi(f))\right| . 
\end{equation}
This gives the relationship between the random fluctuations of the mirror position and the dissipation process. In 1952, this equation was generalized by Callen and Welton \cite{Callen1951} via the famous Fluctuation-Dissipation Theorem which states that if a system dissipates energy, it must also experience noise, and the larger the dissipation, the larger the noise. The PSD of the Langevin force was then generalized to any mechanical system and any dissipation mechanism as: 
\begin{equation}
S_{F_L}(f) = \frac{4k_BT}{\omega}\left|\text{Im}\left(\frac{1}{\chi(f)}\right)\right|, \end{equation}
which is a macroscopic expression of a microscopic phenomena that is responsible for the noise. Now we apply this to compute two dominant thermal noises in the GWD: suspension thermal noise and mirror thermal noise.

\subsubsection{Suspension thermal noise}
The pendulum mirrors in GWDs are placed under high vacuum, which reduces considerably the contribution of the viscous damping. It is observed in this case that the mechanical losses of the mirrors suspensions are dominated by anelastic processes: in the frequencies of interest, each force applied to the suspension results in a motion of constant phase delay. This effect can be incorporated in the harmonic oscillator by adding an imaginary part to the resonance frequency $\omega_0^2\rightarrow\omega_0^2\left(1+i\phi_p\right)$, with $\phi_p$ being the so called loss angle. In this case, the mechanical susceptibility of a single pendulum is:
\begin{equation}
    \chi_p\left(f\right)=\frac{1/m}{\omega_0^2-\omega^2+i\phi_p\omega_0^2} .
\end{equation}To apply the Fluctuation-Dissipation Theorem, one has for $\omega\gg\omega_0$:
\begin{equation}
    \left|\textrm{Im}\left(\chi_p\left(f\right)\right)\right|=\frac{1}{m}\frac{\phi_p\omega_0^2}{\left(\omega_0^2-\omega^2\right)^2+\phi_p^2\omega_0^4}\simeq \frac{1}{m}\frac{\omega_0^2}{\omega^4}\phi_p . 
\end{equation}Now considering 4 suspensions for each of the 4 mirrors of the interferometer, the total suspension thermal noise projected to strain sensitivity is:
\begin{equation}
\label{thermal_suspension}
    S_{h,\textrm{th,sus}}\left(f\right)=\frac{4\times4}{\bar{L}^2}\cdot4\cdot\frac{4k_B T f_0^2}{m\left(2\pi\right)^3f^5}\phi_p .
\end{equation}The recoil force for a suspension is almost entirely gravitational, which is conservative. The only dissipation is related to the residual elastic recoil force through which a small amount of energy is stored. Hence, $\phi_p$ is expected to be very small.  

\subsubsection{Mirror thermal noise}

The mirror (substrate plus coating) being a multiple resonances system, cannot be modelled as a single harmonic oscillator. Instead, each resonance can on its own be considered as a harmonic oscillator. For the sake of simplicity, we assume an anelastic dissipation for all the resonances with a constant loss angle $\phi_m$. Considering only the coordinate along the optical axis $x$ on the mirror surface, the total mechanical susceptibility is:
\begin{equation}
    \chi\left(f\right)=\sum_j\frac{1/m_j}{\omega_j^2-\omega^2+i\phi_m\omega_j^2}=\frac{x_{\textrm{mir}}}{F} , 
\end{equation}where $m_j$ is the mode $j$ mass involving its volume, i.e. the part of the mass actually moving along $x$ when the mode is resonating, $F$ is the amplitude of  a uniform force applied to the mirror surface, and $x_\textrm{mir}$ is the surface displacement of the mirror. The first mirror resonance frequency is much higher than the frequencies of interest, i.e. the detector bandwidth. Hence, for all $j$, $\omega\ll \,\omega_j$:
\begin{eqnarray}
    \nonumber\left|\chi\left(f\right)\right|\simeq\sum_j \frac{1}{m_j \omega_j^2}\quad\textrm{and}\quad \left|\textrm{Im}\left(\chi\left(f\right)\right)\right|\simeq\phi_m \sum_j \frac{1}{m_j \omega_j^2}=\phi_m\left|\chi\left(f\right)\right| ,\\
\end{eqnarray}which are frequency independent down to $f=0$. We can then write the static stored elastic energy when a static force is applied to the mirror as:
\begin{equation}
    W=\frac{1}{2} F_{\textrm{stat}}\times x_\text{mir}\left(f=0\right)=\frac{1}{2} F_\textrm{stat}^2\left|\chi\left(f=0\right)\right|,
\end{equation}
where $F_\textrm{stat}$ is the force $F$ at zero frequency. 
So we have:
\begin{equation}
    \left|\textrm{Im}\left(\chi\left(f\right)\right)\right|=\phi_m\frac{2W}{F_{\textrm{stat}}^2} .
\end{equation}For the sake of simplicity, we assume that the laser beam covers the whole surface area $S_m = \pi r_\textrm{mir}^2$ of a circular mirror with radius $r_\textrm{mir}$. When $F_\textrm{stat}$, which covers the mirror's surface, is applied, the mirror compresses by $\delta \xi\simeq \xi F_\textrm{stat}/ (S_m Y)$, where $\xi$ is the mirror thickness and $Y$ is the Young modulus. Having then $W=\frac{1}{2}F_\textrm{stat}\delta\xi$, we finally obtain  for the mirror's surface displacement :
\begin{equation}
    \left|\textrm{Im}\left(\chi\left(f\right)\right)\right|\simeq\phi_m\times \frac{\xi}{Y S_m} ,\hspace{0.7cm} S_{x,\text{th,mir}}\left(f\right)=\frac{2k_B T}{\pi f}\left|\textrm{Im}\left(\chi\left(f\right)\right)\right| 
\end{equation}Now considering the 4 mirrors, the mirror's thermal noise on the detector is:

\begin{equation}
\label{thermal_mirror}
    S_{h,\textrm{th,mir}}\left(f\right)=\frac{4}{\bar{L}^2}\cdot4\cdot\frac{2k_B T}{\pi f}\cdot\frac{\xi}{Y\pi r_\textrm{mir}^2}\phi_m .
\end{equation}
A more general and accurate approach can be found in \cite{Levin1998}.

\subsection{Quantum noise}
\label{Quantum_noise}
\subsubsection{Quantification of the electromagnetic field}
The classical treatment of quantum noise presented in the earlier sections was useful for an intuitive understanding of the coupling mechanism of shot noise in an interferometer. However, a rigorous and full description of the quantum noise coupling in the interferometer and how to reduce it can only be obtained within a quantum mechanics formalism, that takes into account vacuum fluctuations. For that we will treat each mode of the electromagnetic field as a quantum harmonic oscillator. Due to the scope of this book, our description will be short. A full mathematical description with most of the equations shown in this section can be found in many books of quantum mechanics, and we used as a reference the book \cite{Bachor2019}. 

%
%
%
%

\begin{figure}[h]
\centering
\includegraphics[width=\textwidth]{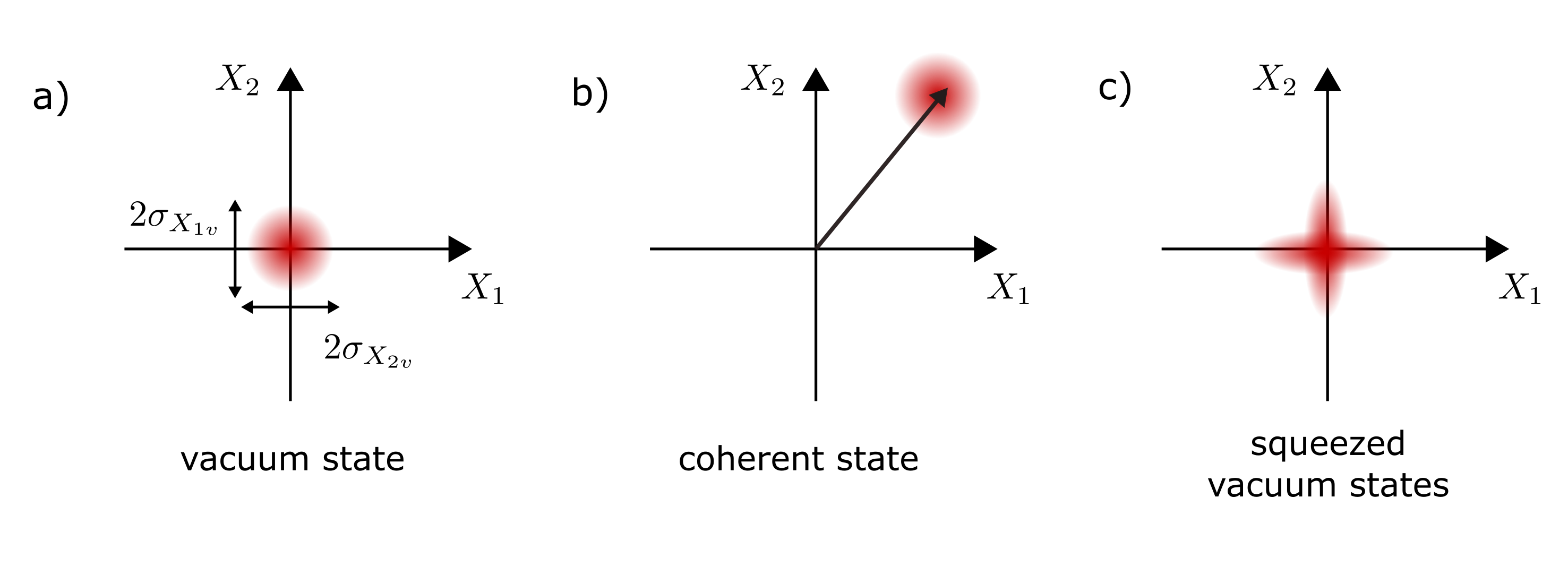}
\caption{Quadrature space representation for a) a vacuum state, b) coherent state, and c) for different squeezed vacuum states.}
\label{fig:quadrature_space}
\end{figure}

In classical physics, a single mode electric field in a plane wave approximation, propagating in vacuum, and with no technical noise, can be decomposed into a sum of two time independent quadratures named $X_1$ and $X_2$ that are oscillating $90^{\circ}$ out of phase with each other: 
\begin{equation}
 \mathcal{E}(t) = \mathcal{E}_0 \left(X_1 \cos \omega_0 t + X_2 \sin \omega_0 t \right) , 
\end{equation} to which corresponds an energy:
\begin{equation}
    E=\kappa\left(\frac{X_1^2}{4}+\frac{X_2^2}{4}\right) ,
\end{equation} 
with $\kappa$ a constant that will be defined later. By analogy with the harmonic oscillator (see Equation \ref{energy_harmonic_oscillator}), one can see that $X_1$ and $X_2$ are the conjugate canonical variables of the position and momentum in a Hamiltonian representation, scaled to be dimensionless. Hence, a similar quantification process can be applied:
\begin{eqnarray}
    &&\left[\hat{X_1},\hat{X_2}\right]=2i , \\
    &&\hat{H}=\kappa\left(\frac{\hat{X}_1^2}{4}+\frac{\hat{X}_2^2}{4}\right)=\kappa\left(\hat{N}+\frac{1}{2}\right) . 
\end{eqnarray}
Two main comments are to be made here:
\begin{itemize}
    \item As for the harmonic oscillator, the single mode electric field is characterized by a set of discrete field eigenstates with energies $E_n=\kappa\left(n+1/2\right)$, $n$ being a positive integer. It can be demonstrated, but it seems obvious to set $\kappa=\hbar\omega_0$ so moving from the state $E_n$ to the state $E_{n+1}$ corresponds to the gain of one photon energy $\hbar \omega_0$.
    \item The vacuum state $\left|n=0\rangle\right.$ corresponds to a non zero energy $E_0=\hbar\omega_0/2$. In the vacuum state (subscript $v$), the quadratures are time dependent and characterized by the quantum variances $\sigma_{\hat{X}_{1v(2v)}}^2$ that are linked with each other by the Heisenberg uncertainty principle:
    \begin{equation}
        \sigma_{\hat{X}_{1v}}\times \sigma_{\hat{X}_{2v}}=1 .
    \end{equation}Since the definition of $X_1$ and $X_2$ is invariant with respect to an arbitrary phase,  $\sigma_{\hat{X}_{1v}}= \sigma_{\hat{X}_{2v}}=1$. In a phasor diagram, the vacuum state is represented by a circle centered at (0,0), like shown in Figure \ref{fig:quadrature_space}.a.
\end{itemize}

\subsubsection{Semi-classical approach : The power quantum fluctuations}
A full quantum treatment of the electric field requires the introduction of the Heisenberg formalism and goes beyond the scope of this book. We can however adopt a semi-classical approach in which the vacuum is represented by a classical fluctuating field of zero mean. A quantum noise limited laser field is called a coherent state \cite{Bachor2019}, and can be formally defined as the action of the displacement operator on the vacuum state in a full quantum mechanics formalism. Such a field is represented in Figure \ref{fig:quadrature_space}.b. The corresponding classical field can be written as a sum of the (already introduced) carrier field (with time independent quadratures) and the vacuum field:
\begin{eqnarray}
   \nonumber \mathcal{E}\left(t\right)=\mathcal{E}_0\left(X_1\cos{\omega_0 t}+X_2\sin{\omega_0 t}\right)+\epsilon_v\left(X_{1v}\left(t\right)\cos{\omega_0 t}+X_{2v}\left(t\right)\sin{\omega_0 t}\right), \label{eq:coherent_field} \\
\end{eqnarray}
where $\epsilon_v$ is the amplitude of the vacuum electric field and $X_{1v(2v)}\left(t\right)$ the corresponding fluctuating quadratures. In order for this classical field to mimic the coherent state behavior, the following conditions must be satisfied:
\begin{itemize}
    \item the average of the quadratures for the vacuum state is zero:
    \begin{equation}
        \langle 0|\hat{X}_{1(2)}|0\rangle = \langle X_{1v(2v)}\left(t\right)\rangle = 0,
    \end{equation}
    \item the variance of the quadratures for the vacuum is equal to one:
    \begin{equation}
        \sigma_{\hat{X}_{1v(2v)}}^2=\langle 0|\hat{X}_{1(2)}^2|0\rangle = \sigma_{X_{1v(2v)}}^2 = \langle X_{1v(2v)}^2\left(t\right)\rangle =1,
    \end{equation}
    \item the vacuum electric field $\epsilon_v$ corresponds to the vacuum energy $\langle 0|\hat{H}|0\rangle=\hbar\omega_0/2$.
\end{itemize}
To the first order of $X_{1v(2v)}$, the power averaged over the light period $2\pi/\omega_0$ is:
\begin{eqnarray}
   \nonumber P\left(t\right)&=&\frac{S}{2\mu_0 c} \mathcal{E}_0^2\left(X_1^2+X_2^2\right)+\frac{S}{\mu_0c}\mathcal{E}_0\epsilon_v\left(X_1X_{1v}\left(t\right)+X_2X_{2v}\left(t\right)\right)\\
   &=& \bar{P}+\delta P_\text{QN}\left(t\right), \label{eq:power_fluctuation}
\end{eqnarray}where $S$ is the beam section and by definition we have $X_1^2+X_2^2=1$. Now, we consider the arbitrary observation time $\Delta t$ required to have for the vacuum energy given by the fluctuating term of Equation \ref{eq:coherent_field}:
\begin{equation}
    \frac{\hbar\omega_0}{2}\simeq \frac{S}{2\mu_0 c}\epsilon_v^2 \Delta t\langle X_{1v}^2+X_{2v}^2\rangle=\frac{S}{\mu_0 c}\epsilon_v^2 \Delta t,
\end{equation}to obtain:
\begin{equation}
    \epsilon_v=\sqrt{\frac{\mu_0c}{2S}\frac{\hbar \omega_0}{\Delta t}}.
\end{equation}
The power fluctuations due to quantum noise $\delta P_\text{QN}$ are then given by Equation \ref{eq:power_fluctuation}:
\begin{equation}
    \delta P_\text{QN}\left(t\right)=\mathcal{E}_0\sqrt{\frac{S}{2\mu_0 c}\frac{\hbar \omega_0}{\Delta t}}\left(X_1X_{1v}\left(t\right)+X_2X_{2v}\left(t\right)\right),
\end{equation}and since $ S_{X_{1v}} (f) = S_{X_{2v}}(f) \equiv S_{X_v}( f)$, the corresponding PSD of power noise due to quantum noise is:
\begin{equation}
    S_\text{QN}\left(f\right)=\hbar \omega_0 \cdot \mathcal{E}_0^2 \frac{S}{2\mu_0 c}\left(X_1^2+X_2^2\right) \cdot \frac{ S_{X_v}\left(f\right)}{\Delta t} =\hbar\omega_0 \bar{P},
\end{equation} which is frequency independent. The second equality is due to the fact that:
\begin{equation}
   \int_{1/\Delta t}S_{X_v}\left(f\right) df=\sigma_{X_v}^2=1 \rightarrow \frac{S_{X_v}\left(f\right)}{\Delta t} =1.
\end{equation}We then obtained the same PSD of the shot noise that was introduced based on a classical model in Section \ref{MI} for shot noise, i.e., $S_\text{QN}\left(f,\bar{P}\right) = S_\text{SN}\left(f,\bar{P}\right)$  and which shows that the vacuum field is at its origin.

\subsubsection{The quantum noise in the interferometer}
\begin{figure}[h]
\centering
\includegraphics[width=\textwidth]{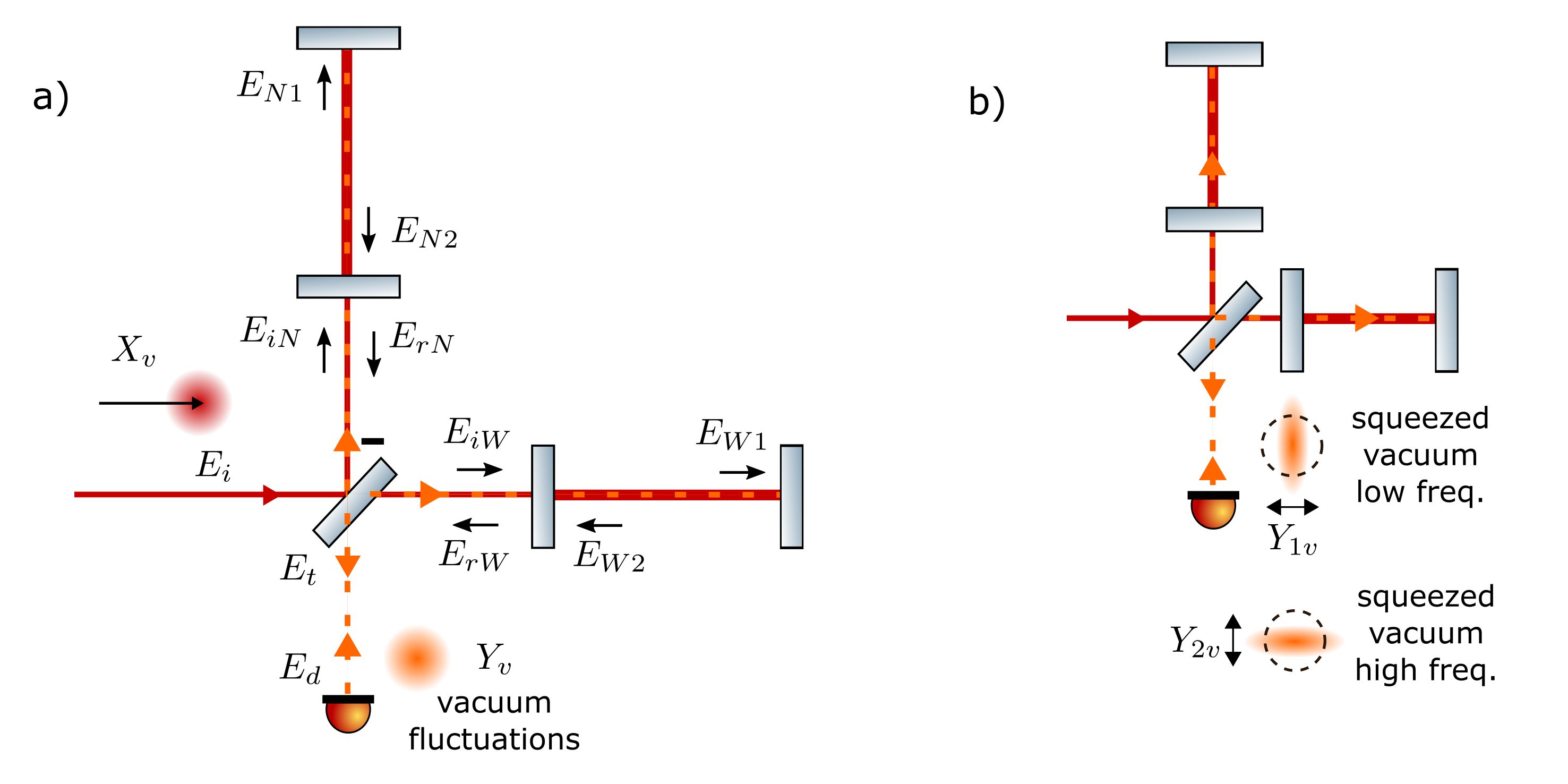}
\caption{a) Vacuum flucuations (dashed orange) coupling from the from the output of a gravitational wave detector, and a coherent field at its input. b) Squeezed vacuum field being injected via the interferometer output in order to reduce its quantum noise.}
\label{fig:vacuum_coupling}
\end{figure}

The vacuum field couples into optical experiments through any channel that is lossy or open. 
The main coupling port of vacuum fluctuations in GWDs happens via the output port of the interferometer \cite{Kimble2001}, which is dark (or close to dark), as shown in Figure \ref{fig:vacuum_coupling}a. As we will now derive, the vacuum fluctuations enters via the interferometer output and are completely (or almost completely) reflected back to the photodetector, resulting in the readout shot noise calculated in Equation \ref{eq:mi_pt_3}. We will also show that the vacuum fluctuations coupling from the input port of the interferometer will be completely (or almost completely) reflected by the interferometer, as well as the laser technical power fluctuations.  

Let us now derive the contribution of quantum noise in the Michelson interferometer shown in Figure \ref{fig:vacuum_coupling}a. For this calculation we will assume the interferometer is operated at the dark fringe and we will ignore technical noise sources. We start by writing the interferometer input field $E_i(t)$, which is a coherent field, as the sum of the carrier and the vacuum fluctuations:
\begin{equation}
E_i(t) = \mathcal{E}_0 \cos(\omega_0 t)+ \epsilon_v \bigl( X_{1v} \cos(\omega_0 t) + X_{2v} \sin(\omega_0 t) \bigr)    .
\end{equation}
For simplicity, we will omit the time dependence on the vacuum quadratures. Similarly, we write the vacuum field coupling at the dark port of the interferometer as:
\begin{equation}
E_d(t) = 
\epsilon_v \bigl( Y_{1v} \cos(\omega_0 t) + Y_{2v} \sin(\omega_0 t) \bigr)    .
\end{equation}
The field directly reflected and transmitted by the beamsplitter at the West and North arms will be given by:
 \begin{align}
E_{iW}(t) =& \frac{1}{\sqrt{2}} \mathcal{E}_0 \cos(\omega_0 t) + \frac{1}{\sqrt{2}} \epsilon_v \bigl( X_{1v} \cos(\omega_0 t) + X_{2v} \sin(\omega_0 t) \bigr) \notag \\ & + \frac{1}{\sqrt{2}} \epsilon_v \bigl( Y_{1v} \cos(\omega_0 t) + Y_{2v} \sin(\omega_0 t) \bigr) \notag \\ & =\frac{1}{\sqrt{2}}\Bigl[\mathcal{E}_0 + \epsilon_v X_{1v} + \epsilon_v Y_{1v}\Bigr] \cos(\omega_0 t)+\frac{1}{\sqrt{2}}\Bigl[\epsilon_v X_{2v} + \epsilon_v Y_{2v}\Bigr] \sin(\omega_0 t) ,   \notag \\
E_{iN}(t) =&  \frac{1}{\sqrt{2}} \Bigl[ -\mathcal{E}_0 - \epsilon_v X_{1v} + \epsilon_v Y_{1v} 
\Bigr] \cos(\omega_0 t) + \frac{1}{\sqrt{2}} \Bigl[ - \epsilon_v X_{2v} + \epsilon_v Y_{2v} 
\Bigr] \sin(\omega_0 t) .
\end{align}   
For the sake of simplicity, we neglect the phase shifts in the small Michelson composed of the beamsplitter and the input cavity mirror with respect to the phase shifts ($\phi_{W,r}$, $\phi_{N,r}$) acquired by the beam in the arm cavities. The reflected fields are then:
\begin{align}
 E_{rW}(t) & = \frac{1}{\sqrt{2}} \Bigl[ \mathcal{E}_0 + \epsilon_v X_{1v}^{(r)} + \epsilon_v Y_{1v}^{(r)} \Bigr] \cos(\omega_0 t + \phi_{W,r}) \notag
\\ & + \frac{1}{\sqrt{2}} \Bigl[ \epsilon_v X_{2v}^{(r)} + \epsilon_v Y_{2v}^{(r)}
\Bigr] \sin(\omega_0 t + \phi_{W,r}), \notag  \\
E_{rN}(t) & =
\frac{1}{\sqrt{2}}
\Bigl[
-\mathcal{E}_0 - \epsilon_v X_{1v}^{(r)} + \epsilon_v Y_{1v}^{(r)}
\Bigr]
\cos(\omega_0 t + \phi_{N,r})
\notag \\ 
& + \frac{1}{\sqrt{2}}
\Bigl[
-\epsilon_v X_{2v}^{(r)} + \epsilon_v Y_{2v}^{(r)}
\Bigr]
\sin(\omega_0 t + \phi_{N,r}) .
\end{align}
Where the superscript $(r)$ stands for the fields in reflection of the optical cavities, in which we will consider that only the phase is changed, not the amplitude. Hence $(X_{1v}(t)^{(r)} \to X_{1v}(t))$. The intracavity fields can be written for $f  \ll f_p$ as:
\begin{align}
E_{{cav,W}}(t)  = &
\sqrt{\frac{\mathcal{F}}{\pi}}
\Bigl[
\mathcal{E}_0 + \epsilon_v X_{1v} + \epsilon_v Y_{1v}
\Bigr]
\cos(\omega_0 t + \phi_{W,cav,i(e)}) \notag \\
& + \sqrt{\frac{\mathcal{F}}{\pi}}
\Bigl[
\epsilon_v X_{2v} + \epsilon_v Y_{2v}
\Bigr]
\sin(\omega_0 t + \phi_{W,cav,i(e)}) ,  \notag  \\
E_{{cav,N}}(t) = &
\sqrt{\frac{\mathcal{F}}{\pi}}
\Bigl[
- \mathcal{E}_0 - \epsilon_v X_{1v} + \epsilon_v Y_{1v}
\Bigr]
\cos(\omega_0 t + \phi_{N,cav,i(e)}) \notag \\
& + \sqrt{\frac{\mathcal{F}}{\pi}}
\Bigl[
-\epsilon_v X_{2v} + \epsilon_v Y_{2v}
\Bigr]
\sin(\omega_0 t + \phi_{N,cav,i(e)}) .
\label{eq:quantum_intracavityfield}
\end{align} where $\phi_{W,cav,i(e)}$ and $\phi_{N,cav,i(e)}$ are the intracavity beams' phases impinging on the input(end) mirrors.
The detected field at the dark port is then:
\begin{align}
E_t = & \frac{1}{\sqrt{2}} E_{rW} + \frac{1}{\sqrt{2}} E_{rN}    \notag \\
= & \frac{1}{2} \, \mathcal{E}_0 \bigl[\cos(\omega_0 t + \phi_{W,r}) - \cos(\omega_0 t + \phi_{N,r})\bigr] \notag \\ & + \frac{1}{2} \, \epsilon_v X_{1v}^{(r)} \bigl[\cos(\omega_0 t + \phi_{W,r}) - \cos(\omega_0 t + \phi_{N,r})\bigr] \notag \\
& + \frac{1}{2} \, \epsilon_v X_{2v}^{(r)} \bigl[\sin(\omega_0 t + \phi_{W,r}) - \sin(\omega_0 t + \phi_{N,r}\bigr]  \notag \\
& + \frac{1}{2} \, \epsilon_v Y_{1v}^{(r)} \bigl[\cos(\omega_0 t + \phi_{W,r}) + \cos(\omega_0 t + \phi_{N,r})\bigr] \notag \\ 
& + \frac{1}{2} \, \epsilon_v Y_{2v}^{(r)} \bigl[\sin(\omega_0 t + \phi_{W,r}) + \sin(\omega_0 t + \phi_{N,r})\bigr] .
\end{align}
As in Equation \ref{eq:common_differential_phases}, we introduce the common and differential phases $\bar{\phi}_r=\left(\phi_{W,r}+\phi_{N,r}\right)/2$ and $\Delta\phi_r=\left(\phi_{W,r}-\phi_{N,r}\right)/2$. The field at the output can be written as: 
\begin{align}
E_t = &
- \mathcal{E}_0
\sin\bigl(\omega_0 t + \bar{\phi}_r\bigr) \sin \Delta \phi_r
- \epsilon_v X_{1v}^{(r)} \sin\bigl(\omega_0 t + \bar{\phi}_r\bigr) \sin \Delta \phi_r \notag \\
&
+ \epsilon_v X_{2v}^{(r)} \cos\bigl(\omega_0 t + \bar{\phi}_r\bigr) \sin \Delta \phi_r     + \epsilon_v Y_{1v}^{(r)} \cos\bigl(\omega_0 t + \bar{\phi}_r\bigr) \cos \Delta \phi_r \notag \\
&
+ \epsilon_v Y_{2v}^{(r)} \sin\bigl(\omega_0 t + \bar{\phi}_r\bigr) \cos \Delta \phi_r ,
\end{align}
which, in the small differential phase approximation can be re-written as:
\begin{equation}
E_t \simeq 
- \mathcal{E}_0 \sin\bigl(\omega_0 t + \bar{\phi}_r\bigr) \sin \Delta \phi_r
+ \epsilon_v \Bigl[
Y_{1v}^{(r)} \cos\bigl(\omega_0 t + \bar{\phi}_r\bigr)
+ Y_{2v}^{(r)} \sin\bigl(\omega_0 t + \bar{\phi}_r\bigr)
\Bigr]    .
\end{equation}
Note that, as expected, the transmitted field depends only on the vacuum field coupling from the interferometer's dark port. The power at the interferometer output will be then given by:
\begin{align}
P_t(t) = & \frac{S}{\mu_0 c} \langle E_t^2(t) \rangle    \notag \\
= & \frac{S}{2 \mu_0 c} \mathcal{E}_0^2 \sin^2 \Delta \phi_r
- \frac{S}{\mu_0 c} \epsilon_v \mathcal{E}_0 Y_{2v}^{(r)}(t) \sin \Delta \phi_r , 
\end{align}
Note that the power fluctuations are solely due to quantum noise, and depends only on the $Y_2$ (phase) quadrature, which is orthogonal to the input carrier at the interferometer. This coupling is also known as quantum readout noise, since the noise couples at the same quadrature of the signal expected from a GW. Their power spectral density can be calculated as:
\begin{equation}
S_{P_t} (f)=
\left[
\frac{S^2}{\mu_0^2 c^2} \, \epsilon_v^2 \, \mathcal{E}_0^2 \, \sin^2 \Delta\phi_r
\right] S_{Y_{2v}^{(r)}}
= \hbar \omega_0 \, \bar{P}_t    ,
\end{equation}
which gives the same result as in Equation \ref{eq:mi_pt_3}. This quadrature $Y_{2v}$ entirely couples to the output port of the detector and not to the input port. Hence the power recycling just enhances the input power by the factor $G$ and one gets the same Equation \ref{Shot}.  The strain sensitivity limited by the quantum readout noise is then: 
\begin{equation}
S_{h,\textrm{QN}}\left(f\right) = \frac{h_p c \lambda_0}{64\mathcal{F}^2G\bar{P_0} \bar{L}^2}\left(1 + \frac{f^2}{f_p^2}\right) .
\label{eq:strain_qn}
\end{equation}

The vacuum fluctuations will also result in the so-called quantum radiation pressure noise: the fluctuating power impinging on each suspended mirror applies a fluctuating force of quantum origin. The circulating power in the west and north arm cavities can be calculated from Equations \ref{eq:quantum_intracavityfield} as:
\begin{align}
& P_{\mathrm{cav}W} (t) \simeq
\frac{S}{2 \mu_0 c} \frac{\mathcal{F}}{\pi}
\left(
\mathcal{E}_0^2 + 2 \mathcal{E}_0  \epsilon_v(X_{1v}(t) + Y_{1v}(t))
\right)   ,  \\ 
& P_{\mathrm{cav}N}  (t) \simeq
\frac{S}{2 \mu_0 c} \frac{\mathcal{F}}{\pi}
\left(
\mathcal{E}_0^2 + 2 \mathcal{E}_0  \epsilon_v (X_{1v}(t) - Y_{1v}(t)) 
\right) ,
\end{align}
which depends only on the quadrature of the vacuum fields that is aligned with the carrier (amplitude quadrature). The arm cavities length change due to quantum radiation pressure can be obtained from the equation of motion of a free falling mirror (and no damping) with a driving radiation pressure force $F_\textrm{rp}(t) = 2 P(t)/c$:
\begin{equation}
\delta L_{W(N)}(t) = \frac{4}{mc}  \int_{t'} \int_t P_{\mathrm{cav}W(N)}(t) dt dt',      
\end{equation}
where and additional factor of 2 was inserted to account for the radiation pressure effect in both cavitiy mirrors. The differential displacement and its PSD can then be calculated:
\begin{align}
   & \delta L_\textrm{qrp} (t) = \delta L_{W}(t) - \delta L_{N}(t) = -\frac{8 S\mathcal{F}}{mc \mu_0 \pi} \mathcal{E}_0  \epsilon_v \int_{t'} \int_t Y_{1v}(t) dt dt',\\
   &  S_{\delta L_\textrm{qrp}}(f) = \frac{64 S^2 \mathcal{F}^2 }{(mc \mu_0 \pi)^2} \frac{1}{(2\pi f)^4}\mathcal{E}_0 ^2 \epsilon_v^2 \cdot S_{Y_{1v}} (f).
\end{align}
From these equations, one can see that only the amplitude quadrature of the vacuum field coupling at the dark port $Y_{1v}$ will contribute to the differential displacement since the contribution from $X_{1v}$ results in common motion in the interferometer's arms. The strain sensitivity limited by quantum radiation pressure noise is then:
\begin{equation}
S_{h,\textrm{qrp}}(f) =  \frac{4}{\bar{L}^2}\frac{4\mathcal{F}^2}{m^2\pi^6}\frac{h_P c}{\lambda_0}\frac{G\bar{P}_0}{f^4}     
\label{eq:strain_qrp}
\end{equation}

Since quantum noise has its origin the vacuum, one can redistribute the noise uncertainty via a vacuum squeezing process \cite{Bachor2019}. This results in the two different ellipses illustrated on Figure \ref{fig:quadrature_space}.c. There one can see that the noise circle has been squeezed in one direction, resulting in a quantum state in which the uncertainty in one quadrature is reduced at the penalty of increasing the uncertainty in the orthogonal quadrature.

GWDs inject squeezed vacuum states at their output port (see Figure \ref{fig:vacuum_coupling}c) to improve their sensitivity \cite{Ganapathy2023}. They inject the so called frequency dependent squeezing in which the quadrature that is squeezed depends on the frequency. At low frequencies, where the detector is limited by quantum radiation pressure noise, the squeezed quadrature is the one aligned with the input interferometer carrier (amplitude quadrature), i.e., $Y_{1v}$. At high frequencies, where the sensitivity is limited by the quantum readout noise, the squeezed quadrature is the one orthogonal to the input carrier, i.e., $Y_{2v}$.  

\section{Conclusion: full sensitivity of a GWD}
Now that we calculated the contribution of the main noise sources to the interferometer output, we can finally compute the total interferometer strain sensitivity curve $S_{h,\textrm{tot}}\left(f\right)$. The total contribution from differential length noise can be calculated from  seismic noise, suspension thermal noise, mirror thermal noise, and quantum radiation pressure noise (Equations \ref{seismic}, \ref{thermal_suspension}, \ref{thermal_mirror} and \ref{eq:strain_qrp}):
    \begin{equation}
        S_{h,\delta L_-}\left(f\right) = S_{h,\text{sis}}\left(f\right)+S_{h,\textrm{th,sus}}\left(f\right)+S_{h,\textrm{th,mir}}\left(f\right)+S_{h,\text{qrp}}\left(f\right).
    \end{equation}
Hence, the total sensitivity will be given by:
    \begin{equation}
    \label{sensitivity_h}
        S_{h,\textrm{tot}}\left(f\right)=S_{h,\textrm{QN}}\left(f\right)+S_{h,\delta L_-}\left(f\right)  .
    \end{equation}
\begin{figure}[h]
\centering
\includegraphics[width=0.8\textwidth]{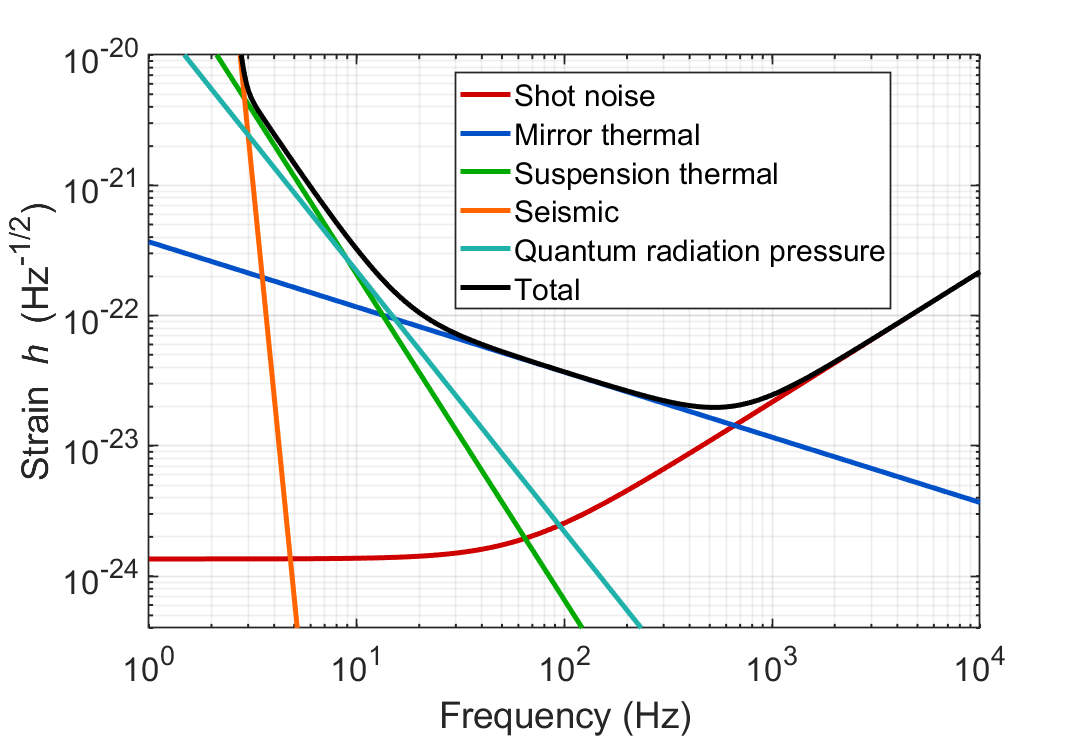}
\caption{ASD of the total strain sensitivity of a gravitational wave detector, obtained from the simplified theoretical model developed in this book, which nevertheless is close to the sensitivity of real detectors. The noise contribution for different noise sources is shown separately. The minimum strain sensitivity is 10$^\text{-23}\,\text{Hz}^\text{-1/2}$ at $\simeq$ 200 Hz.}
\label{fig:sensitivity_curve}
\end{figure}
As an example, we will plot the sensitivity considering current parameters of the Virgo GWD: input power of $\bar{P}_0=25\,\textrm{W}$, an arm length of $\bar{L} = 3\,\textrm{km}$, a finesse for the arm cavities of $\mathcal{F}\simeq 400$, and the power recycling gain of $G\simeq 50$. We considered that the vertical seismic displacement noise is the same as the typical horizontal seismic displacement noise from Equation \ref{eq:horizontal_seismic}. For the thermal noise, we consider the loss angle for the pendulum and the mirror substrate to be $\phi_p\simeq10^{-9}$ and $\phi_m\simeq10^{-6}$. Finally, each mirror have a mass of $m=40\,\textrm{kg}$. The corresponding sensitivity curve is displayed in Figure \ref{fig:sensitivity_curve} by the black curve, together with the individual contributions of the different noise sources. Even though based on simplified models, the result is quite similar to measured sensitivity curves that one can find for example in \cite{AdV}. From the figure, it is evident that the detector sensitivity is limited by different noise sources at different frequencies:
\begin{itemize}
    \item seismic noise for $f\lesssim 5\,\textrm{Hz}$,
    \item suspension thermal noise for $5\,\textrm{Hz}\lesssim f\lesssim 10\,\textrm{Hz}$ ,
    \item mirror thermal noise for $10\,\textrm{Hz}\lesssim f\lesssim 200\,\textrm{Hz}$ which, given the quality of the substrates, is mainly due to the coatings, and
    \item quantum readout noise for $f\gtrsim 200\,\textrm{Hz}$.
\end{itemize}
In current detectors, low-frequency sensitivity is in reality limited by noise from different control loops, which are not described here. In the following, we provide a non-exhaustive list of upgrades aimed at improving the sensitivity of future detectors:
\begin{itemize}
    \item lowering the resonance frequency $f_0$ of the suspensions in order to shift both seismic noise and suspension thermal noise towards low frequencies. The options are, however, limited, since $f_0$ scales with the square root of the suspension length, 
    \item increasing the input power to reduce the quantum readout noise contribution. This, however, enhances the contribution of quantum radiation pressure noise, whose ASD scales as $\sqrt{\bar{P}_0}$ (see Equation \ref{eq:strain_qrp}). To mitigate this drawback, squeezing techniques have been proposed and are already implemented in current detectors \cite{Ganapathy2023}, and
    \item reducing thermal noise by operating at cryogenic temperatures. While promising, this approach faces significant technical challenges \cite{Ushiba2021}.  
\end{itemize}


\end{document}